\documentclass{amsart}
\usepackage[latin1]{inputenc}
\usepackage{amsmath}
\usepackage{amsthm}
\usepackage{amsfonts}
\usepackage{amssymb}
\newtheorem{theorem}{Theorem}[section]
\newtheorem{lemma}[theorem]{Lemma}
\newtheorem{proposition}[theorem]{Proposition}
\newtheorem{definition}[theorem]{Definition}

\newtheorem{remark}[theorem]{Remark}
\newtheorem{corollary}[theorem]{Corollary}
\newtheorem{example}[theorem]{Example}

\newcommand{\supp}{\mathop{\mathrm{supp}}}
\newcommand{\vp}{\mathop{\mathrm{vp}}}

\newcommand{\diag}{\mathop{\mathrm{diag}}}

\begin{document}
\title[On generalized Lorentz Geometry]{On Lorentz geometry in algebras of generalized functions}
\author[E. Mayerhofer]{Eberhard Mayerhofer}
\address{University of Vienna, Faculty of Mathematics, Nordbergstrasse 15, 1090 Vienna, Austria}
\thanks{Work supported by FWF research grants P16742-N04 and Y237-N13}
\email{eberhard.mayerhofer@univie.ac.at}  \maketitle
\begin{abstract}
We introduce a concept of causality in the framework of generalized
pseudo-Riemannian Geometry in the sense of J.F. Colombeau and establish the inverse Cauchy-Schwarz inequality in this context. As an application, we prove a dominant energy
condition for some energy tensors as put forward in Hawking and
Ellis's book \lq\lq The large scale structure of space-time\rq\rq. Our work is based
on a new characterization of free elements in finite dimensional modules over the ring of generalized numbers.
\end{abstract}
\section{Introduction}
The theory of distributions is an indispensable tool for investigating linear partial differential equations. As an example
we mention the theorem of Malgrange-Ehrenpreis which asserts that every linear PDE with constant coefficients has a fundamental solution in $\mathcal D'$. However, there are natural limitations
in its applicability to non-linear problems. Concerning the analysis of PDEs with non-constant coefficients,
the desire to solve a differential equation in all of $\mathcal D'$ soon requires the
definition of products of distributions. Such definitions, however, are usually restricted to specific subspaces of $\mathcal D'$
(e.g.\ Sobolev spaces) or fail to display certain algebraic properties of a product. More explicitly, we mention
\begin{itemize}
\item (lack of consistency) The definitions vary from application to application, e.g., 
the definition $H\delta=c\delta$ may be reasonable for every complex number $c$
(cf.\ \cite{Bible}, Examples 1.1.1).
\item (product properties) The product in general lacks nice properties, such as commutativity or associativity.  Indeed, assuming we are given an associative product $\circ$ on $\mathcal D'$ and let $\vp(1/x)$ denote the principal value of $1/x$. Then we
would have
\[
\delta=\delta\circ(x \circ \vp(1/x))=(\delta\circ x)\circ
\vp(1/x)=0,
\]
which is impossible, since $\delta\neq 0$. For a more detailed study we refer to (\cite{MObook}).
\end{itemize}
The need for defining an unrestricted multiplication of elements of $\mathcal D'$ therefore motivates the search for non-linear extension of the space of distribution.
More precisely it is desirable to have an associative, commutative algebra $({\mathcal G},+,\circ)$ such that:
\begin{enumerate}
\item\label{eig1alg} There exists a linear embedding $\iota:{\mathcal D}' \hookrightarrow {\mathcal G}$ such that
$\iota(1)$ is the unit in ${\mathcal G}$.
\item There exist derivation operators $D_i: {\mathcal G} \to
{\mathcal G}$ ($1\leq i \leq s$), which are linear and satisfy the
Leibniz-rule.
\item $D_i \mid_{{\mathcal D}'} = \frac{\partial}{\partial x_i}$
($1\le i \le s$), that is the derivation operators restricted to
$\mathcal D'$ are the usual partial derivations.
\item \label{propfour} $\circ \mid_{{\mathcal C}^\infty\times
{\mathcal C}^\infty}$ is the point-wise product of
functions.
\end{enumerate}
Item (\ref{propfour}) corresponds to the natural requirement that the new
product should coincide with the usual point-wise product on a
\lq\lq reasonable\rq\rq\ subspace of $\mathcal D'$. Schwartz's celebrated impossibility result (\cite{Schw1}) states that
such an algebra does not exist if (\ref{propfour}) is weakened to the respective requirement
on $C^k$ functions (the space of $k$-times differentiable functions).

The construction of a differential algebra $(\mathcal
G,+,\circ)$ which satisfies (\ref{eig1alg})--(\ref{propfour}) was achieved by J.F. Colombeau (\cite{Colombeau,C}). The key idea of
his construction is regularization of distributions. Generalized functions are basically 
described by nets of smooth functions parametrized by the smoothing parameter and satisfying
a specific asymptotic growth property with respect to the latter. Now there are
a number of such algebras of generalized functions. For a general construction scheme,
cf. \cite{Bible}. 

A non-linear theory of generalized functions in a geometric setting has
been developed by Kunzinger and Steinbauer (\cite{PenroseKS, KS1}, cf.\ also \cite{Bible} and section \ref{sec2}). This approach allows for mathematically rigorous investigations of distributional geometries. In the context of general relativity this theory has proved valuable for formulating and solving problems, e.g. concerning weak singularities such as cosmic strings and impulsive gravitational waves (cf.\ \cite{CVW, HS}). These are singularities which admit a locally bounded metric, but the curvature has to be calculated on the distributional level. Since the curvature tensor is a non-linear function of the metric tensor and its first two derivatives, its calculation in general involves ill-defined products of distributions, unless one deviates from the distributional framework. 

This paper is a result of recent research on the intersection of general relativity and the theory of generalized function algebras. Related work concerns, for instance, classifying singularities of space-times following a concept of C.J.S. Clarke: Singularities in space-times are considered essential if they disrupt the evolution of the wave-equation (\lq\lq generalized hyperbolicity\rq\rq\ , cf.\ \cite{Clarke, VW}). It turned out that for a deeper understanding of singular space-times as modelled in algebras of generalized functions it is indispensable to reinterpret the notion of causality in this framework. The present article 
meets this requirement by contributing some algebraic foundations for ongoing research in this field. In addition, we hope that the results laid out in this paper may also be of independent interest to the field of nonlinear generalized functions.
\subsection*{Program of the paper}
In sections \ref{sec2} and \ref{secin} we recall constructions in generalized
pseudo-Riemannian geometry and we revisit invertibility and positivity issues
in the special algebra.
Sections \ref{sec4}, \ref{sec5} and \ref{sec6} form the core of the paper. Section \ref{sec4} 
deals with symmetric generalized matrices, introducing a notion of generalized eigenvalues of the latter.
By means of the positivity concept revisited in section \ref{secin} we introduce
in section \ref{sec5} a generalized concept of causality. Furthermore, the inverse Cauchy-Schwarz inequality is proved in this context. As
an application a dominant energy condition for a class of generalized Energy tensors is established.
The final section \ref{sec6} presents generalized point value characterizations of generalized pseudo-Riemannian metrics
and of causality of generalized vector fields by means of the theory developed in the preceding two sections.
The paper ends with an appendix on further algebraic properties of finite dimensional modules over the ring of generalized numbers.
\section{Preliminaries}\label{sec2}
\subsection{The ring of generalized numbers and a partial order}
Throughout  the symbol $\mathbb K$ denotes $\mathbb R$ resp.\ $\mathbb C$. In what follows we use the index set $I:=(0,1]\subset \mathbb R$. We shall construct generalized numbers based on \lq\lq nets of real or complex number\rq\rq\  $(a_\varepsilon)_\varepsilon$ indexed by $\varepsilon\in I$, i.e., on elements of $\mathbb K^I$.
The ring of generalized numbers over $\mathbb K$ is constructed in the following way: Given the ring of
moderate nets of numbers 
\[
\mathcal E_M:=\{(x_{\varepsilon})_{\varepsilon}\in\mathbb K^I \mid \exists\;
m:\vert
x_{\varepsilon}\vert=O(\varepsilon^m)\,(\varepsilon\rightarrow 0)\}
\]
and, similarly, the ideal of negligible nets in $\mathcal E(\mathbb
K)$ which are of the form 
\[
\mathcal N:=\{(x_{\varepsilon})_{\varepsilon}\in\mathbb K^I\mid \forall\;
m:\vert
x_{\varepsilon}\vert=O(\varepsilon^m)\,(\varepsilon\rightarrow
0)\},
\]
we may define the generalized numbers as the factor ring
\[
\widetilde{\mathbb K}:=\mathcal E_M/\mathcal N.
\]
Given a moderate net $(a_\varepsilon)_\varepsilon\in\mathcal E_M$, we denote by
$[(a_\varepsilon)_\varepsilon]$ its class in $\widetilde{\mathbb K}$.
Next we show how a partial order $\leq$ can be introduced on
$\widetilde{\mathbb R}$ (cf.\ \cite{MOHor,PSMO}). For $a\,,b\in\widetilde{\mathbb R}$, we say $a\leq b$
if and only if there exist representatives $(a_\varepsilon)_\varepsilon$, $(b_\varepsilon)_\varepsilon$
of $a,b$ such that for each $\varepsilon>0$ we have $a_\varepsilon\leq b_\varepsilon$, in the usual order
on the real line. This is equivalent to saying that
for arbitrary representatives $(\bar a_\varepsilon)_\varepsilon$, $(\bar b_\varepsilon)_\varepsilon$
there is a negligible number $(n_\varepsilon)_\varepsilon$ such that
\[
(\forall \varepsilon>0)(\bar a_\varepsilon\leq \bar b_\varepsilon+n_\varepsilon).
\]
$(\widetilde{\mathbb R},\, \leq)$ is a partially ordered ring (cf.\ \cite{Bible}, Proposition 1.2.36), however
$\leq$ is not a total order on $\widetilde{\mathbb R}$. As an example for a pair of numbers
which are not comparable with respect to this order, we define $c,\,d\in\widetilde{\mathbb R}$ on the level of representatives
by
\[
c_\varepsilon:=\begin{cases} 1, \textit{ if } \varepsilon =1/n\; (n\in\mathbb N)\\ 0, \textit{ otherwise}\end{cases},\;\;\; d_\varepsilon:=1-c_\varepsilon,\; (\varepsilon\in I).
\]
Another difference to the situation on the real numbers is the following. Suppose we are given a number $a\in\widetilde{\mathbb R}$, with a representatives 
$(a_\varepsilon)_\varepsilon$ satisfying 
\begin{equation}\label{posdef}
\forall \varepsilon>0, a_\varepsilon>0.
\end{equation}
This does not imply that $a$ is invertible: Note that even $0$ admits positive representatives, for instance
$n_\varepsilon:=\exp(-1/\varepsilon)$. Indeed, $(n_\varepsilon)_\varepsilon$ tends to zero faster than any power of $\varepsilon$, for $\varepsilon\rightarrow 0$, hence
is a representative of $0$.

The above example motivates us to introduce, apart from $\geq 0$, a further order. We will call an element $a\in\widetilde{\mathbb R}$ strictly
positive if and only if $a$ admits a representative $(a_{\varepsilon})_{\varepsilon}$ such that 
\begin{equation}\label{strposdef}
(\exists\, m\geq 0)(\exists\, \varepsilon_0)(\forall\,\varepsilon<\varepsilon_0,\, a_{\varepsilon}\geq \varepsilon^m)
\end{equation}
In this case we shall write $a>0$. Contrary to the above situation (\ref{posdef}) where $a\geq 0$, strict positivity implies invertibility. The main reason for this difference is that property (\ref{strposdef}) is stable under a change of representatives, whereas (\ref{posdef}) is not. For more information on positivity as well as on invertibility we refer to section \ref{secin}.

Let $A\subset I$, then the characteristic function $\chi_A\in\widetilde{\mathbb R}$ is given by the class of $(\chi_{\varepsilon})_{\varepsilon}$, where
\[
\chi_{\varepsilon}:=\begin{cases} 1,\qquad\mbox{if}\qquad \varepsilon\in A\\ 0, \qquad\mbox{otherwise}\end{cases}.
\]
$\widetilde{\mathbb R}^n$ shall be considered as an $\widetilde{\mathbb R}$--module of dimension $n\geq 1$.
Clearly the latter can also be constructed by a quotient of \lq\lq moderate nets of vectors\rq\rq\ by \lq\lq negligible nets of vectors\rq\rq.
\subsection{The special Colombeau algebra on manifolds}
This section is devoted to introducing the special algebra on manifolds in a coordinate
independent way as in \cite{K}. A translation into coordinate expressions of the respective objects is
given in the end of this section.

The material presented until the end of section \ref{sec2} stems from the original sources
\cite{K, KS}. For a comprehensive presentation we refer to the--meanwhile standard reference on generalized function algebras
-- \cite{Bible}. Moreover, for further works in geometry based
on Colombeau's ideas we refer to (\cite{GlobTh,KKo1,KU2,3MikesV,KS,KSV,GenConKSV}).

In this paper, $X$ shall denote a paracompact,
smooth Hausdorff manifold of dimension $n$ and by $\mathcal P(X)$ we
denote the space of linear differential operators on $X$. $K\subset\subset X$ denotes a set $K$ compactly contained in $X$.
The special algebra of generalized functions on $X$ is constructed as
the quotient $\mathcal G(X):=\mathcal E_M(X)/\mathcal N(X)$, where
the ring of moderate (resp.\ negligible) nets of smooth functions is given by
\begin{align}\nonumber
\mathcal E_M(X):=\{(u_{\varepsilon})_{\varepsilon}\in (C^{\infty}(X))^I\mid\forall\;K\subset\subset X\;\forall\;P\in\mathcal P(X)\;\exists\;N\in\mathbb N:\\\sup_{x\in K}\vert Pu_{\varepsilon}\vert=O(\varepsilon^{-N})\,(\varepsilon\rightarrow 0)\}
\end{align}
resp.\
\begin{align}\nonumber
\mathcal N(X):=\{(u_{\varepsilon})_{\varepsilon}\in (C^{\infty}(X))^I\mid\forall\;K\subset\subset X\;\forall\;P\in\mathcal P(X)\;\forall\;m\in\mathbb N:\\\sup_{x\in K}\vert Pu_{\varepsilon}\vert=O(\varepsilon^m)\,(\varepsilon\rightarrow 0)\}.
\end{align}
Given a moderate net $(u_\varepsilon)_\varepsilon\in \mathcal E_M(X)$ we dennote by $[(u_\varepsilon)_\varepsilon]$ its class in
$\mathcal G(X)$.

The $C^{\infty}$-sections of a vector bundle $(E,X,\pi)$ with base
space $X$ we denote by $\Gamma(X, E)$. Moreover, let $\mathcal
P(X,E)$ be the space of linear partial differential operators acting
on $\Gamma(X,E)$. The $\mathcal G(X)$-module of generalized
sections $\Gamma_{\mathcal G}(X,E)$ of a vector bundle
$(E,X,\pi)$ on $X$ is defined similarly as (the algebra of
generalized functions on $X$) above, in that we use asymptotic
estimates with respect to the norm induced on the respective fibers by some arbitrary
Riemannian metric. That is, we define the
quotient
\[
\Gamma_{\mathcal G}(X,E):=\Gamma_{\mathcal E_M}(X,E)/\Gamma_{\mathcal N}(X,E),
\]
where the module of moderate (resp.\ negligible) nets of sections is given by
\begin{align}\nonumber
\Gamma_{\mathcal E_M}(X,E):=\{(u_{\varepsilon})_{\varepsilon}\in
(\Gamma(X,E))^I\mid\forall\;K\subset\subset X\;\forall\;P\in\mathcal
P(X, E)\;\exists\;N\in\mathbb N:\\\sup_{x\in
K}\|Pu_{\varepsilon}\|=O(\varepsilon^N)\,(\varepsilon\rightarrow 0)\}
\end{align}
resp.\
\begin{align}\nonumber
\Gamma_{\mathcal N}(X,E):=\{(u_{\varepsilon})_{\varepsilon}\in
(\Gamma(X,E))^I\mid\forall\;K\subset\subset X\;\forall\;P\in\mathcal
P(X, E)\;\forall\;m\in\mathbb N:\\\sup_{x\in
K}\|Pu_{\varepsilon}\|=O(\varepsilon^m)\,(\varepsilon\rightarrow 0)\}.
\end{align}
In this article we shall deal with generalized sections of the tensor
bundle $\mathcal T^{r}_{s}(X)$ over $X$, which we denote by
\[
\mathcal G^{r}_{s}(X):=\Gamma_{\mathcal G}(X,\mathcal T^{r}_{s}(X)).
\]
We call elements of $\mathcal G^{r}_{s}(X)$ {\it generalized tensors of type
$(r,s)$}. 
We end this section by translating the global description of generalized vector bundles
into coordinate expressions. Following the notation of \cite{KS}, we denote by
$(V,\Psi)$ a vector bundle chart over a chart $(V,\psi)$ of the base $X$. With $\mathbb R^{n'}$, the typical fibre,
we can write:
\[
\Psi:\pi^{-1}(V)\rightarrow\psi(V)\times \mathbb R^{n'},
\]
\[
z\mapsto(\psi(p),\psi^1(z),\dots,\psi^ {n'}(z)).
\]
Let now $s\in\Gamma_{\mathcal G}(X,E)$. Then the local expressions of $s$, $s^i=\Psi^i\circ s\circ \psi^{-1}$ lie
in $\mathcal G(\psi(V))$. 

An equivalent \lq\lq local definition\rq\rq\  of generalized vector bundles can be achieved by 
defining moderate nets $(s_\varepsilon)_\varepsilon$ of smooth sections $s_\varepsilon$
to be such for which the local expressions $s_\varepsilon^i=\Psi^i\circ s_\varepsilon\circ \psi^{-1}$
are moderate, that is $(s_\varepsilon^i)_\varepsilon\in\mathcal E_M(\psi(V))$ (the notion negligible is defined completely analogously).
This follows from the fact that every linear differential operator can be localized  (cf.\ \cite{Bible}, p. 289).
\subsection{Uniqueness in $\mathcal G(X)$}\label{uniqueness}
A function $f\in\mathcal G(X)$ can be evaluated on standard points $x\in X$. To be more precise, let
$(f_\varepsilon)_\varepsilon$ be a representative of $f$. Then the mapping
\begin{equation}\label{eqxx}
f: X\rightarrow \widetilde{\mathbb R},\qquad x\mapsto f(x):=(f_\varepsilon(x))_\varepsilon+\mathcal N
\end{equation}
is well defined (cf.\ \cite{MO1}). It is customary to call $f(x)$ the point value of $f$ at $x$. Note that the above constitutes a slight abuse of notation: On the one hand, $f$ is a generalized function and on the other hand, $f$ denotes the evaluation mapping (\ref{eqxx}). 

Generalized functions are {\it  not} uniquely determined by evaluation on standard points (\cite{Eyb4, MO1}). To illustrate this important feature of generalized function algebras
we recall Example 2.1 from \cite{MO1}:

Take some $\varphi\geq 0\in\mathcal D(\mathbb R)$ with $\supp\varphi\in[-1,1]$ and $\int \varphi=1$ and set $u_\varepsilon:=\varphi_\varepsilon(x-\varepsilon)$,
where $\varphi_\varepsilon(y):=\frac{1}{\varepsilon}\varphi(\frac{y}{\varepsilon})$. Then $(u_\varepsilon)_\varepsilon\in\mathcal E_M(\mathbb R)$, so
$u:=[(u_\varepsilon)_\varepsilon]\in\mathcal G(\mathbb R)$. One can easily see that for all $x\in\mathbb R$, $u_\varepsilon(x)=0$, whenever $\varepsilon$
is sufficiently large. Hence, $u(x)=0$ in $\widetilde{\mathbb R}$. But $u\neq 0$.

However, if we allow the point $x$ to vary with $\varepsilon$ (on the level of representatives this means inserting a net $(x_\varepsilon)_\varepsilon$ into $(f_\varepsilon)_\varepsilon$ instead of standard points only as in eq.\ (\ref{eqxx})), we can uniquely determine generalized functions by evaluation. More precisely, the following holds
(\cite{MO1}, Theorem 2.4 and \cite{KS}, Theorem 1):
\begin{theorem}
Let $f\in\mathcal G(X)$. The following are equivalent:
\begin{enumerate}
 \item $f=0$ in $\mathcal G(X)$,
 \item $f(x_c)=0$ in $\widetilde{\mathbb R}$ for each $x_c\in\widetilde{X}_c$.
\end{enumerate}
\end{theorem}
Here $\widetilde X_c$ denotes the class of nets $(x_\varepsilon)_\varepsilon$ of compactly supported points
factored by the equivalence relation $\sim$ given by
\[
(x_\varepsilon)_\varepsilon\sim (y_\varepsilon)_\varepsilon\Leftrightarrow \forall m\geq 0:\; d(x_\varepsilon,y_\varepsilon)=O (\varepsilon^m), \;\textit{whenever }\;(\varepsilon\rightarrow 0)
\]
where $d$ is the distance function induced by an arbitrary Riemannian metric. Completely analogous to (\ref{eqxx}), the evaluation
of $f$ at points in $\widetilde{X}_c$ is well defined. 

\subsection{Generalized pseudo-Riemannian metrics}\label{introducerepseudoriemannereetconnexione}
We begin by recalling the following characterization of non-degenerateness of symmetric (generalized) tensor fields of type (0,2) on $X$ (\cite{KS1}, Theorem 3.1)
\begin{theorem}\label{chartens02}
Let $g\in \mathcal G^0_2(X)$. The following are equivalent:
\begin{enumerate}
\item \label{chartens021} For each chart $(V_{\alpha},\psi_{\alpha})$ and each $\widetilde x\in (\psi_{\alpha}(V_{\alpha}))^{\sim}_c$ the map\\
$g_{\alpha}(\widetilde x): \widetilde{\mathbb R}^n\times \widetilde{\mathbb R}^n\rightarrow \widetilde{\mathbb R}$ is symmetric and non-degenerate.
\item $g: \mathcal G^0_1(X)\times \mathcal G^0_1(X)\rightarrow \mathcal G(X)$ is symmetric and for each chart $(V_{\alpha},\psi_{\alpha})$, $\det g_\alpha$ is invertible in $\mathcal G(\psi_\alpha(V_\alpha))$.
\item \label{chartens023} For each chart $(V_{\alpha},\psi_{\alpha})$, $\det g_\alpha$ is invertible in $\mathcal G(\psi_\alpha(V_\alpha))$ and for each relatively compact open set $V\subset X$ there exists a representative $(g_{\varepsilon})_{\varepsilon}$ of $g$ and $\varepsilon_0>0$ such that $g_{\varepsilon}\mid_V$ is a smooth pseudo-Riemannian metric for all $\varepsilon<\varepsilon_0$.
\end{enumerate}
\end{theorem}
Furthermore, the index of $g\in \mathcal G^0_2(X)$ is introduced in the following well defined way
(cf.\ Definition 3.2 and Proposition 3.3 in \cite{KS1}):
\begin{definition} \label{defpseud}
Let $g\in \mathcal G^0_2(X)$ satisfy one (hence all) of the equivalent conditions in Theorem \ref{chartens02}. If there exists some $j\in\mathbb N$ with the property that for each relatively compact open set $V\subset X$ there exists a representative $(g_{\varepsilon})_{\varepsilon}$ of $g$ as in Theorem \ref{chartens02} (\ref{chartens023}) such for each $\varepsilon<\varepsilon_0$ the index of $g_{\varepsilon}$ is equals $j$ we say $g$ has index $j$. Such symmetric 2-forms we call generalized pseudo-Riemannian metrics on $X$.
\end{definition}
The field of generalized pseudo--Riemannian geometry deals with pairs $(X, g)$, where $g$ is a pseudo--Riemannian metrics on $X$ with index
$\nu$. 
\section[Invertibility revisited]{Invertibility and strict positivity in generalized function algebras revisited}\label{secin}
This section is devoted to elaborating a new characterization of invertibility as well as of strict positivity of generalized numbers resp.\ functions. The first investigation on which many works in this field are based was done by M.\ Kunzinger and R.\ Steinbauer in \cite{KS1}; the authors of the latter work
established the fact that invertible generalized numbers are precisely such for which the modulus of any representative is bounded from below by a fixed power of the smoothing parameter (cf.\ the proposition below). It is, however, noteworthy that component-wise invertibility on the level of representatives describes invertibility of generalized numbers entirely:
\begin{proposition}\label{genpointinv}
Let $\gamma\in\widetilde{\mathbb R}$. The following are equivalent:
\begin{enumerate}
\item \label{inv1} $\gamma$ is invertible.
\item \label{inv2} $\gamma$ is strictly nonzero, that is: for some (hence any) representative $(\gamma_{\varepsilon})_{\varepsilon}$ of $\gamma$ there exists an $m_0$ and an $\varepsilon_0\in I$ such that for each $\varepsilon<\varepsilon_0$ we have
$\vert \gamma_{\varepsilon}\vert>\varepsilon^{m_0}$.
\item \label{inv3} For each representative $(\gamma_{\varepsilon})_{\varepsilon}$ of $\gamma$ there exists some $\varepsilon_0\in I$ such that for all
$\varepsilon<\varepsilon_0$ we have $\gamma_{\varepsilon}\neq 0$.
\item \label{inv4} $\vert \gamma\vert$ is strictly positive.
\end{enumerate}
\end{proposition}
\begin{proof}
Since (\ref{inv1}) $\Leftrightarrow$ (\ref{inv2}) by (\cite{KS1}, Theorem 1.2.38) and (\ref{inv1}) $\Leftrightarrow $ (\ref{inv4}) follows from the definition of
strict positivity, we only need to establish the equivalence (\ref{inv2}) $\Leftrightarrow$ (\ref{inv3}) in order to complete
proof.
As the reader can easily verify, the definition of strictly non-zero is independent of the representative, that is for each representative $(\gamma_{\varepsilon})_{\varepsilon}$ of $\gamma$ we have
some $m_0$ and some $\varepsilon_0$ such that for all $\varepsilon<\varepsilon_0$ we have $\vert\gamma_{\varepsilon}\vert>\varepsilon^{m_0}$. By this consideration (\ref{inv3}) follows from (\ref{inv2}).
In order to show the converse direction, we proceed by an indirect argument. Assume there exists a representative $(\gamma_{\varepsilon})_{\varepsilon}$
of $\gamma$ such that for some zero sequence $\varepsilon_k\rightarrow 0$ ($k\rightarrow \infty$) we have $\vert\gamma_{\varepsilon_k}\vert<\varepsilon_k^k$ for each $k>0$.
Define a moderate net $(\hat{\gamma}_{\varepsilon})_{\varepsilon}$ in the following way:
\[
\hat{\gamma}_{\varepsilon}:=\begin{cases} 0\qquad\mbox{if}\qquad \varepsilon=\varepsilon_k\\ \gamma_{\varepsilon} \qquad\mbox{otherwise} \end{cases}.
\]
It can then easily be seen that $(\hat{\gamma}_{\varepsilon})_{\varepsilon}-(\gamma_{\varepsilon})_{\varepsilon}\in \mathcal N(\mathbb R)$
which means that $(\hat{\gamma}_{\varepsilon})_{\varepsilon}$ is a representative of $\gamma$ as well. However the latter violates
(\ref{inv3})  and we are done.
\end{proof}
We can characterize the strict order relation on the ring of generalized real numbers in a similar manner:
\begin{proposition}\label{secinverweis}
Let $\gamma\in\widetilde{\mathbb R}$. The following are equivalent:
\begin{enumerate}
\item \label{inv11} $\gamma$ is strictly positive, that is: for some (hence any) representative $(\gamma_{\varepsilon})_{\varepsilon}$ of $\gamma$ there exists an $m_0$ and an $\varepsilon_0\in I$ such that for each $\varepsilon<\varepsilon_0$ we have
$\gamma_{\varepsilon}>\varepsilon^{m_0}$.

\item \label{inv21} $\gamma$ is strictly nonzero and has a representative $(\gamma_{\varepsilon})_{\varepsilon}$ which is positive for each index $\varepsilon>0$.
\item \label{inv31} For each representative $(\gamma_{\varepsilon})_{\varepsilon}$ of $\gamma$ there exists some $\varepsilon_0\in I$ such that for all
$\varepsilon<\varepsilon_0$ we have $\gamma_{\varepsilon}> 0$.
\end{enumerate}
\end{proposition}
The statement can be shown similarly to the preceding one. 

Next, we draw our attention to the question of invertibility and strict positivity of generalized functions. We start with the definition
of the latter:
\begin{definition}
A function $f\in\mathcal G(X)$ is called strictly positive in $\mathcal G(X)$, if $f$ is invertible and if
for each compact subset $K\subset X$ there
exists a representative $(f_{\varepsilon})_{\varepsilon}$ of $f$ which is non-negative on $K$. We shall write $f>0$. $f\in\mathcal G(X)$ is called
strictly negative in $\mathcal G(X)$, if $-f>0$ on $X$.
\end{definition}
Next, we show that  Propositions \ref{genpointinv} and \ref{secinverweis} have immediate generalizations to generalized functions on $X$	:
\begin{theorem}\label{downhilliseasier}
Let $u\in\mathcal G(X)$. The following  are equivalent:
\begin{enumerate}
\item \label{invf1} $u$ is invertible (resp.\ strictly positive).
\item \label{invf1extra} For each compactly supported point $x_c\in \widetilde{X}_c$, $u(x_c)$ is an invertible element of
$\widetilde{\mathbb R}$.
\item \label{invf2} For each representative $(u_{\varepsilon})_{\varepsilon}$ of $u$ and each compact set $K$ in $X$ there exists some
$\varepsilon_0\in I$ and some $m_0$ such that for all $\varepsilon<\varepsilon_0$ we have $\inf_{x\in K} \vert u_{\varepsilon}(x)\vert>\varepsilon^{m_0}$ (resp.\ $\inf_{x\in K} u_{\varepsilon}(x)>\varepsilon^{m_0}$).
\item  \label{invf3} For each representative $(u_{\varepsilon})_{\varepsilon}$ of $u$ and each compact set $K$ in $X$ there exists some
$\varepsilon_0\in I$ such that $\forall\; x\in K\;\forall\; \varepsilon<\varepsilon_0: u_{\varepsilon}(x)\neq 0$ (resp.\ $u_{\varepsilon}(x)>0$).
\end{enumerate}
\end{theorem}
\begin{proof}
We only show that the characterization of invertibility holds, the rest of the statement is then clear. (\ref{invf1})$\Leftrightarrow$(\ref{invf2}) hold according to (\cite{KS1}, Proposition 2.1). Furthermore the equivalence (\ref{invf1}) $\Leftrightarrow$ (\ref{invf1extra}) can be easily elaborated by modifying suitably the proofs of
Theorem 2.4 in (\cite{MO1}) resp.\ of Proposition 3.4 in (\cite{PSMO}). It is therefore sufficient to establish the equivalence
of the latter two statements. Since (\ref{invf2})$\Rightarrow$(\ref{invf3}) is evident, we finish the proof by showing the converse direction.
Assume (\ref{invf2}) does not hold, then there exists a compactly supported sequence $(x_k)_k\in X^{\mathbb N}$ such that for some representative $(u_{\varepsilon})_{\varepsilon}$ of $u$ we have $\vert u_{\varepsilon_k}(x_k)\vert<\varepsilon_k^k$
for each $k$. Similarly to the proof of Proposition \ref{genpointinv} we observe that $(\hat u_{\varepsilon})_{\varepsilon}$ defined by
\[
\hat{u}_{\varepsilon}:=\begin{cases} u_{\varepsilon}-u_{\varepsilon}(x_k),\qquad\mbox{if}\qquad \varepsilon=\varepsilon_k\\ u_{\varepsilon}, \qquad\mbox{otherwise}\end{cases}
\]
yields another representative of $u$ which, however, violates (\ref{invf3}) and we are done.
\end{proof}
We will frequently employ the notion of positivity characterized here (e.g.\ in Definition \ref{causaldef1} of causality).
However, positivity in the generalized sense is a fundamental property which has proved useful in other contexts as well. We refer here to papers by Oberguggenberger et al 
concerning positivity and positive definiteness in generalized function algebras (\cite{MOHor}) and also on elliptic regularity for partial differential equations
with generalized coefficients (\cite{PSMO}).
\section{Matrices over $\widetilde{\mathbb R}$}\label{sec4}
We denote by $\widetilde{\mathbb R}^{n^2}:=\mathcal M_n(\widetilde{\mathbb R})$ the ring of
 $n\times n$ matrices over $\widetilde{\mathbb R}$. A matrix $A$ is called orthogonal, if $UU^t=\mathbb I$ in $\widetilde{\mathbb R}^{n^2}$ and $\det U=1$ in $\widetilde{\mathbb R}$.
Clearly, there are two different ways to introduce $\widetilde{\mathbb R}^{n^2}$:
\begin{remark}\rm
Denote by $\mathcal E_M(\mathcal M_n(\mathbb R))$ the ring of moderate nets of
$n\times n$ matrices over $\mathbb R$, a subring of $\mathcal M_n(\mathbb R)^I$. Similarly let $\mathcal N(\mathcal M_n(\mathbb R))$
denote the ideal of negligible nets of real $n\times n$ matrices. There is a ring isomorphism
$\varphi: \widetilde{\mathbb R}^{n^2}\rightarrow \mathcal E_M(\mathcal M_n(\mathbb R))/\mathcal N(\mathcal M_n(\mathbb R))$.
\end{remark}
For the convenience of the reader we repeat Lemma 2.6 from \cite{KS1}:
\begin{lemma}\label{nondeg}
Let $A\in\widetilde{\mathbb R}^{n^2}$. The following are equivalent:
\begin{enumerate}\label {charnondeg}
\item \label{charnondeg1} $A$ is non-degenerate, that is, $\xi \in \widetilde{\mathbb R}^n,\;\xi^t A\eta=0$ for each $\eta\in\widetilde{\mathbb R}^n$ implies $\xi=0$.
\item \label{charnondeg2} $A: \widetilde{\mathbb R}^n\rightarrow \widetilde{\mathbb R}^n$ is injective.
\item \label{charnondeg3} $A: \widetilde{\mathbb R}^n\rightarrow \widetilde{\mathbb R}^n$ is bijective.
\item\label{charnondeg4} $\det A$ is invertible in $\widetilde{\mathbb R}$.
\end{enumerate}
\end{lemma}
Note that the equivalence of (\ref{charnondeg1})--(\ref{charnondeg3}) and (\ref{charnondeg4}) results from the fact that in $\widetilde{\mathbb R}$ any nonzero non-invertible element is a zero-divisor. Since we deal with symmetric matrices throughout, we start by giving a basic
characterization of symmetry of generalized matrices:
\begin{lemma}\label{symmetry}
Let $A\in\widetilde{\mathbb R}^{n^2}$. The following are equivalent:

\begin{enumerate}
\item \label{asymmetry} $A$ is symmetric, that is $A=A^t$ in $\widetilde{\mathbb R}^{n^2}$.
\item \label{bsymmetry} There exists a symmetric representative $(A_{\varepsilon})_{\varepsilon}:=((a_{ij}^{\varepsilon})_{ij})_{\varepsilon}$ of $A$.
\end{enumerate}
\end{lemma}
\begin{proof}
Since (\ref{bsymmetry}) $\Rightarrow$ (\ref{asymmetry}) is clear, we only need to show (\ref{asymmetry}) $\Rightarrow$ (\ref{bsymmetry}). Let $((\bar a_{ij}^{\varepsilon})_{ij})_{\varepsilon}$ a representative of $A$. Symmetrizing yields the desired representative
\[
(a_{ij}^{\varepsilon})_{\varepsilon}:=\frac{ (\bar a_{ij}^{\varepsilon})_{\varepsilon}+(\bar a_{ji}^{\varepsilon})_{\varepsilon} } {2}
\]
of $A$. This follows from the fact that for each pair $(i,j)\in\{1,\dots,n\}^2$ of indices one has $(\bar a_{ij}^{\varepsilon})_{\varepsilon}-(\bar a_{ji}^{\varepsilon})_{\varepsilon} \in \mathcal N(\mathbb R)$ due to the symmetry of $A$.
\end{proof}
Denote by $\|\,\|_F$ the Frobenius norm on $\mathcal M_n(\mathbb C)$.
In order to prepare a notion of eigenvalues for symmetric matrices, we repeat a numeric result given in \cite{SJ} (Theorem 5.\ 2):
\begin{theorem}\label{perturbation}
Let $A\in\mathcal M_n(\mathbb C)$ be a Hermitian matrix with eigenvalues $\lambda_1\geq\dots\geq\lambda_n$. Denote by
$\widetilde A$ a non-Hermitian perturbation of $A$, i.\ e., $E=\widetilde A- A$ is not Hermitian. We further call the eigenvalues of $\widetilde A$ (which might be complex) $\mu_k+i\nu_k\;(1\leq k\leq n)$ where $\mu_1\geq\dots\geq\mu_n$. In this notation, we have
\[
\sqrt{ \sum_{k=1}^n\vert(\mu_k+i\nu_k)-\lambda_k \vert^2 }\leq\sqrt 2\| E\|_F.
\]
\end{theorem}
\begin{definition}\label{eigenvalues}
Let $A\in\widetilde{\mathbb R}^{n^2}$ be a symmetric matrix and let $(A_{\varepsilon})_{\varepsilon}$ be an arbitrary representative of $A$. Let for any $\varepsilon\in I$, $\theta_{k,\varepsilon}:=\mu_{k,\varepsilon}+i\nu_{k,\varepsilon}\;(1\leq k\leq n)$ be the eigenvalues of $A_{\varepsilon}$ ordered by the size of the real parts, i.\ e., $\mu_{1,\varepsilon}\geq\dots\geq\mu_{n,\varepsilon}$.
The generalized eigenvalues $\theta_k\in\widetilde{\mathbb C}\;(1\leq k\leq n)$ of $A$ are defined as the classes $(\theta_{k,\varepsilon})_{\varepsilon}+\mathcal N(\mathbb C)$.
\end{definition}
\begin{lemma}\label{schur}
Let $A\in\widetilde{\mathbb R}^{n^2}$ be a symmetric matrix. Then the eigenvalues $\lambda_k\;(1\leq k\leq n)$ of $A$ as introduced in Definition
\ref{eigenvalues} are well defined elements of $\widetilde{\mathbb R}$. Furthermore, there exists an orthogonal $U\in \widetilde{\mathbb R}^{n^2}$ such that
\begin{equation}\label{eqdecschur}
U A U^t=\diag (\lambda_1,\dots,\lambda_n).
\end{equation}
We call $\lambda_i\;(1\leq i\leq n)$ the eigenvalues of $A$. $A$ is non-degenerate if and only if all generalized eigenvalues are invertible.
\end{lemma}
Before we prove the lemma, we note that throughout this paper we shall omit the term \lq\lq generalized\rq\rq\  (eigenvalues) and we shall call the generalized numbers constructed in the above way simply \lq\lq eigenvalues\rq\rq\ (of a generalized symmetric matrix).
\begin{proof}
Due to Lemma \ref{symmetry} we may choose a symmetric representative $(A_{\varepsilon})_{\varepsilon}=((a_{ij}^{\varepsilon})_{ij})_{\varepsilon}\in\mathcal E_M(\mathcal M_n(\mathbb R))$ of $A$ . For any $\varepsilon$, denote by $\lambda_{1,\varepsilon}\geq\dots\geq\lambda_{n,\varepsilon}$ the resp.\ (real) eigenvalues
of $(a_{ij}^{\varepsilon})_{ij}$ ordered by size. For any $i\in\{1,\dots,n\}$, define $\lambda_i:=(\lambda_{i,\varepsilon})_\varepsilon+\mathcal N(\mathbb R)\in\widetilde{\mathbb R}$. For the well-definedness of the eigenvalues of $A$, we only need to show that for any other (not necessarily symmetric) representative of $A$, the resp.\ net of eigenvalues lies in the same class of $\mathcal E_M(\mathbb C)$; note that the use of complex numbers is indispensable here. Let $
(\widetilde A_{\varepsilon})_{\varepsilon}=((\widetilde a_{ij}^{\varepsilon})_{ij})_{\varepsilon}$ be another representative of $A$. Denote by $\mu_{k,\varepsilon}+i\nu_{k+\varepsilon}$ the eigenvalues of $\widetilde A_{\varepsilon}$ for any $\varepsilon\in I$ such that
the real parts are ordered by size, i.\ e., $\mu_{1,\varepsilon}\geq\dots\geq \mu_{n,\varepsilon}$.
Denote by $(E_{\varepsilon})_{\varepsilon}:=(\widetilde A_{\varepsilon})_{\varepsilon}-(A_{\varepsilon})_{\varepsilon}$.
Due to Theorem \ref{perturbation} we have for each $\varepsilon\in I$:
\begin{equation}\label{nullboundsE}
\sqrt{\sum_{k=1}^n\vert(\mu_{k,\varepsilon}+i\nu_{k,\varepsilon})-\lambda_{k,\varepsilon}\vert^2 }\leq\sqrt 2\|E_{\varepsilon}\|_F.
\end{equation}
Since $(E_{\varepsilon})_{\varepsilon}\in\mathcal N(\mathcal M_n(\mathbb R))$, (\ref{nullboundsE}) implies for any $k\in\{1,\dots,n\}$ and any $m$,
\[
\vert(\mu_{k,\varepsilon}+i\nu_{k,\varepsilon})-\lambda_{k,\varepsilon}\vert=O(\varepsilon^m)\;(\varepsilon\rightarrow 0)
\]
which means that the resp.\ eigenvalues of $(A_\varepsilon)_\varepsilon$ and of $(\widetilde A_\varepsilon)_\varepsilon$ in the above order belong to the same class in $\mathcal E_M(\mathbb C)$. In particular they yield the same elements of $\widetilde{\mathbb R}$.
The preceding argument and Lemma \ref{symmetry} show that without loss of generality we may construct the eigenvalues of $A$ by means of a symmetric representative $(A_{\varepsilon})_{\varepsilon}=((a_{ij}^{\varepsilon})_{ij})_{\varepsilon}\in\mathcal E_M(\mathcal M_n(\mathbb R))$. For such a choice we have for any $\varepsilon$
an orthogonal matrix $U_{\varepsilon}$ such that
\[
U_{\varepsilon}A_{\varepsilon}U_{\varepsilon}^t=\diag(\lambda_{1,\varepsilon},\dots,\lambda_{n,\varepsilon}),\;\lambda_{1,\varepsilon}\geq\dots\geq\lambda_{n,\varepsilon}.
\]
Declaring $U$ as the class of $(U_{\varepsilon})_{\varepsilon}\in\mathcal E_M(\mathcal M_n(\mathbb R))$ yields the proof of the second claim,
since orthogonality for any $U_{\varepsilon}$  implies orthogonality of $U$ in $\mathcal M_n(\widetilde{\mathbb R})$. Finally, decomposition (\ref{eqdecschur}) gives, by applying the multiplication theorem for determinants and the orthogonality of $U$, $\det A=\prod_{i=1}^n \lambda_i$. This shows in conjunction with Lemma \ref{nondeg} that invertibility of all eigenvalues is a sufficient and necessary condition for the non-degenerateness of $A$ and we are done.
\end{proof}
\begin{remark}\rm
A remark on the notion eigenvalue of a generalized symmetric matrix $A\in\widetilde{\mathbb R}^{n^2}$ is in order: Since for any eigenvalue $\lambda$ of $A$
we have $\det (A-\lambda \mathbb I)=\det (U(A-\lambda \mathbb I)U^t)=\det ((UAU^t)-\lambda \mathbb I)=0$, Lemma \ref{charnondeg}
implies that $A-\lambda \mathbb I: \widetilde{\mathbb R}^n\rightarrow\widetilde{\mathbb R}^n$ is not injective. However, again by the same lemma, $\det (A-\lambda \mathbb I)=0$ is not necessary for $A-\lambda \mathbb I$ to be not injective, and a $\theta\in\widetilde{\mathbb R}$ for which $A-\theta I$ is not injective
need not be an eigenvalue of $A$. More explicitly, we give two examples of possible scenarios here:
\begin{enumerate}
\item  Let $\forall\;i\in\{1,\dots,n\}: \lambda_i\neq 0$ and for some $i$ let $\lambda_i$ be a zero divisor. Then besides
$A-\lambda_i\;(i=1,\dots,n)$, also $ A: \widetilde{\mathbb R}^n\rightarrow\widetilde{\mathbb R}^n$ fails to be injective.
\item  \lq\lq Mixing\rq\rq\ representatives of $\lambda_i,\lambda_j\; (i\neq j)$ might give rise to generalized numbers $\theta\in\widetilde{\mathbb R}, \,\theta\neq \lambda_j\,\forall j\in\{1,\dots,n\}$ for
which $A-\theta \mathbb I$ is not injective as well. Consider for the sake of simplicity the matrix $D:=\diag (1,-1)\in \mathcal M_2(\mathbb R)$.
A rotation $U_{\varphi}:=\left(\begin{array}{cc} \cos(\varphi)& \sin(\varphi)\\-\sin(\varphi)&\cos(\varphi) \end{array}\right)$ yields
by matrix multiplication
\[
U_{\varphi} D U_{\varphi}^t=\left(\begin{array}{cc} \cos(2\varphi)& -\sin(2\varphi)\\-\sin(2\varphi)&-\cos(2\varphi) \end{array}\right).
\]
\end{enumerate}
The choice of $\varphi=\pi/2$ therefore switches the order of the entries of $D$, that is $U_{\pi/2}DU_{\pi/2}^t=\diag(-1,1)$.
Define $U,\lambda$ as the classes of $(U_{\varepsilon})_{\varepsilon}, (\lambda_{\varepsilon})_{\varepsilon}$ defined by
\[
U_{\varepsilon}:=\begin{cases} I:\;\varepsilon\in I\cap\mathbb Q\\ U_{\pi/2}:\;\mbox{else} \end{cases},
\]
\[
\lambda_{\varepsilon}:=\begin{cases} 1:\;\varepsilon\in I\cap\mathbb Q\\ -1\;\mbox{else} \end{cases},
\]
further define $\mu\in\widetilde{\mathbb R} $ by $\mu+\lambda=0$.
Then we have for $A:= [(D)_{\varepsilon}]$:
\[
UDU^t=\diag(\lambda,\mu).
\]
\end{remark}
Therefore as shown above, $D-\lambda \mathbb I,\; D-\mu \mathbb I$ are not injective considered as maps $\widetilde{\mathbb R}^n\rightarrow\widetilde{\mathbb R}^n$.
But neither $\lambda$, nor $\mu$ are eigenvalues of $D$.
\begin{definition}\label{indexmatrixdef}
Let $A\in\widetilde{\mathbb R}^{n^2}$. We denote by $\nu_{+}(A)$ (resp.\ $\nu_{-}(A)$) the number of
strictly positive (resp.\ strictly negative) eigenvalues, counting multiplicity. Furthermore, if $\nu_{+}(A)+\nu_{-}(A)=n$,
we simply write $\nu(A):=\nu_{-}(A)$. If $A$ is symmetric and $\nu(A)=0$, we call $A$ a positive definite symmetric matrix. If $A$ is symmetric and $\nu_{+}(A)+\nu_{-}(A)=n$ and $\nu(A)=1$, we say $A$ is a symmetric $L$-matrix.
\end{definition}
The following corollary shows that for a symmetric non-degenerate matrix in $\widetilde{\mathbb R}^{n^2}$ counting $n$ strictly positive resp.\ negative eigenvalues is equivalent to having a (symmetric) representative for which any $\varepsilon$-component has the same number (total $n$) of positive resp.\ negative real eigenvalues. The proof can be obtained by using ideas of the proof of Proposition \ref{secinverweis}.
\begin{corollary}\label{charindexmatrix}
Let $A\in\widetilde{\mathbb R}^{n^2}$ be symmetric and non-degenerate and $j\in\{1,\dots,n\}$. The following are equivalent:
\begin{enumerate}
\item $\nu_{+}(A)+\nu_{-}(A)=n$, $\nu(A)=j$.
\item \label{charindexmatrix2} For each symmetric representative $(A_{\varepsilon})_{\varepsilon}$ of $A$ there exists some $\varepsilon_0 \in I$ such that for any $\varepsilon<\varepsilon_0$ we have for the eigenvalues
$\lambda_{1,\varepsilon}\geq\dots\geq\lambda_{n,\varepsilon}$  of $A_{\varepsilon}$:
\[
\lambda_{1,\varepsilon},\dots,\lambda_{n-j,\varepsilon}>0,\;\;\lambda_{n-j+1,\varepsilon},\dots,\lambda_{n,\varepsilon}<0.
\]
\end{enumerate}
\end{corollary}
\section{Causality and the inverse Cauchy-Schwarz inequality}\label{sec5}
In a free module over a commutative ring $R\neq \{0\}$, any two bases have the same cardinality. Therefore, any free module $\mathfrak M_n$ of dimension $n\geq 1$ (i.\ e., with a basis having $n$ elements) is isomorphic to $R^n$ considered as module over $R$ (which is free, since it has the canonical basis). As a consequence we may confine ourselves to considering the module $\widetilde{\mathbb R}^n$ over $\widetilde{\mathbb R}$ and its submodules. We further assume that from now on $n$, the dimension of $\widetilde{\mathbb R}^n$, is greater than $1$. It is quite natural to start with an appropriate version of the Steinitz exchange lemma:
\begin{proposition}\label{steinitzprop}
Let $\mathcal B=\{v_1,\dots,v_n\}$ be a basis for $\widetilde{\mathbb R}^n$. Let $w=\lambda_1 v_1+\dots+\lambda_nv_n\in \widetilde{\mathbb R}^n$ such that for some $j\;(1\leq j\leq n)$,
$\lambda_j$ is strictly nonzero. Then, also $\mathcal B':=\{v_1,\dots,v_{j-1},w,v_{j+1},\dots,v_n\}$ is a basis for $\widetilde{\mathbb R}^n$.
\end{proposition}
\begin{proof}
Since every strictly nonzero number is invertible, one can prove the claim by using the proof of the well known one in the vector space setting.
\end{proof}
\begin{definition}\label{bilformdef}
Let $b:\widetilde{\mathbb R}^n\times\widetilde{\mathbb R}^n\rightarrow \widetilde{\mathbb R}$ be a symmetric bilinear form. Suppose there exists a number $j\in\mathbb N_0$ such that for some basis $\mathcal B:=\{e_1,\dots,e_n\}$ of $\widetilde{\mathbb R}^n$ we have $\nu((b(e_i,e_j))_{ij})=j$. Then we call $j$ the index of $b$. If $j=0$ we say that $b$ is positive definite and if $j=1$ we call $b$ a symmetric bilinear form of Lorentzian signature.
\end{definition}
Note that as in the classical setting, there is no notion of 'eigenvalues' of a symmetric bilinear form, since a change of coordinates that is not induced by an orthogonal matrix need not conserve the eigenvalues of the original coefficient matrix.
We are obliged to show that the notion above is well defined. The main argument is Sylvester's inertia law (cf.\ \cite{GF}, pp.\ 306):
\begin{proposition}
The index of a bilinear form $b$ on $\widetilde{\mathbb R}^n$ as introduced in Definition \ref{bilformdef} is well defined.
\end{proposition}
\begin{proof}
Let $\mathcal B$, $\mathcal B'$ be bases of $\widetilde{\mathbb R}^n$ and let $A$ be a matrix describing a linear map
which maps $\mathcal B$ onto $\mathcal B'$ (this map is uniquely determined in the sense that it only depends on the order of the basis vectors of the resp.\ bases). Let $B$ be the coefficient matrix of the given bilinear form $b$ and let
further $k:=\nu(B)$. The change of bases
results in a 'generalized' equivalence transformation of the form
\[
B\mapsto T:=A^tBA,
\]
$T$ being the coefficient matrix of $h$ with respect to $\mathcal B'$. We only need to show that $\nu(B)=\nu(T)$. Since the index of a matrix is well defined (and this again follows from Lemma \ref{schur}, where it is proved that the eigenvalues of a symmetric generalized matrix are well defined), it is sufficient to show that for one any symmetric representative $(T_{\varepsilon})_{\varepsilon}$ of $T$ there exists an $\varepsilon_0\in  I$ such that for each $\varepsilon<\varepsilon_0$ we have
\[
 \lambda_{1,\varepsilon}>0,\dots,\lambda_{n-k,\varepsilon}>0,\lambda_{n-k+1,\varepsilon}<0,\dots,\lambda_{n-k,\varepsilon}<0,
\]
where $(\lambda_{i,\varepsilon})_{\varepsilon}$ ($i=1,\dots,n$) are the ordered eigenvalues of $(T_{\varepsilon})_{\varepsilon}$. To this end, let $(B_{\varepsilon})_{\varepsilon}$ be a symmetric representative
of $B$, and define by $(T_{\varepsilon})_{\varepsilon}$ a representative of $T$ component-wise via
\[
T_{\varepsilon}:=A_{\varepsilon}^tB_{\varepsilon}A_{\varepsilon}.
\]
Clearly $(T_{\varepsilon})_{\varepsilon}$ is symmetric. For each $\varepsilon$ let
$\lambda_{1,\varepsilon}\geq\dots\geq\lambda_{n,\varepsilon}$ be the ordered eigenvalues of $T_{\varepsilon}$
and let $\mu_{1,\varepsilon}\geq\dots\geq\mu_{n,\varepsilon}$ be the ordered eigenvalues of $B_{\varepsilon}$.
Since $A$ and $B$ are non-degenerate, there exists some $\varepsilon_0\in I$ and an integer $m_0$ such that
for each $\varepsilon<\varepsilon_0$ and for each $i=1,\dots,n$ we have
\[
\vert \lambda_{i,\varepsilon}\vert\geq \varepsilon^{m_0}\qquad\mbox{and}\qquad \vert \mu_{i,\varepsilon}\vert\geq \varepsilon^{m_0}.
\]
Furthermore due to our assumption $k=\nu(B)$, therefore taking into account the component-wise order of the eigenvalues
$\mu_{i,\varepsilon}$, for each $\varepsilon<\varepsilon_0$ we have:
\[
\mu_{i,\varepsilon}\geq \varepsilon^{m_0}\;\;(i=1,\dots,n-k)\qquad \mbox{and}\qquad \mu_{i,\varepsilon}\leq -\varepsilon^{m_0}\;\;(i=n-k+1,\dots,n).
\]
As a consequence of Sylvester's inertia law we therefore have for each $\varepsilon<\varepsilon_0$:
\[
\lambda_{i,\varepsilon}\geq \varepsilon^{m_0}\;\; (i=1,\dots,n-k)\qquad \mbox{and}\quad \lambda_{i,\varepsilon}\leq -\varepsilon^{m_0}\;\; (i=n-k+1,\dots,n),
\]
since for each $\varepsilon<\varepsilon_0$ the number of positive resp.\ negative eigenvalues of $B_{\varepsilon}$
resp.\ $T_{\varepsilon}$ coincides. We have thereby shown that $\nu(T)=k$ and we are done.
\end{proof}
\begin{definition}
Let $b:\widetilde{\mathbb R}^n\times\widetilde{\mathbb R}^n\rightarrow \widetilde{\mathbb R}$ be a symmetric bilinear form. A basis $\mathcal B:=\{e_1,\dots,e_k\}$ of $\widetilde{\mathbb R}^n$ is called an orthogonal basis with respect to $b$, if $b(e_i,e_j)=0$ whenever $i\neq j$.
\end{definition}
\begin{corollary}\label{exorthbasis}
Any symmetric bilinear form $b$ on $\widetilde{\mathbb R}^n$ admits an orthogonal basis.
\end{corollary}
\begin{proof}
Let $\mathcal B:=\{v_1,\dots,v_n\}$ be some basis of $\widetilde{\mathbb R}^n$, then the coefficient matrix $A:=(b(v_i,v_j))_{ij}\in\widetilde{\mathbb R}^{n^2}$ is symmetric. Due to Lemma \ref{schur}, there is an orthogonal matrix $U\in\widetilde{\mathbb R}^{n^2}$ and generalized numbers
$\theta_i\;(1\leq i\leq n)$ (the so-called eigenvalues) such that $UAU^t=\diag( \theta_1,\dots,\theta_n)$. Therefore the (clearly non-degenerate) matrix $U$ induces a mapping $\widetilde{\mathbb R}^n\rightarrow\widetilde{\mathbb R}^n$ which maps $\mathcal B$ onto some basis $\mathcal B'$
which is orthogonal.
\end{proof}
\begin{definition}
Let $\lambda_1,\dots,\lambda_k\in\widetilde{\mathbb R}$ ($k\geq 1$). Then the span of $\lambda_i$, $(1\leq i\leq k)$ in $\widetilde{\mathbb R}^n$ is denoted by $\langle\{\lambda_1,\dots,\lambda_n\}\rangle$.
\end{definition}
We now introduce a notion of causality in our framework:
\begin{definition}\label{causaldef1}
Let $g$ be a symmetric bilinear form of Lorentzian signature on $\widetilde{\mathbb R}^n$. Then we call $u\in\widetilde{\mathbb R}^n$
\begin{enumerate}
\item time-like, if $g(u,u)<0$,
\item null, if $u=0$ or $u$ is free and $g(u,u)=0$,
\item space-like, if $g(u,u)>0$.
\end{enumerate}
Furthermore, we say two time-like vectors $u,v$ have the same time-orientation whenever $g(u,v)<0$.
\end{definition}
Note that there exist elements in $\widetilde{\mathbb R}^n$ which are neither time-like, nor null, nor space-like.

The next statement provides a crucial characterization of free elements in $\widetilde{\mathbb R}^n$. We shall repeatedly make use of it in the sequel.
\begin{theorem}\label{freechar}
Let $v$ be an element of $\widetilde{\mathbb R}^n$. Then the following are equivalent:
\begin{enumerate}
\item\label{freechar1} For any positive definite symmetric bilinear form $h$ on $\widetilde{\mathbb R}^n$ we have
\[
h(v,v)>0
\]
\item\label{freechar2} The coefficients of $v$ with respect to some (hence any) basis span $\widetilde{\mathbb R}$.
\item \label{freechar3}$v$ is free.
\item \label{freechar4} The coefficients $v^i$ ($i=1,\dots,n$) of $v$ with respect to some (hence any) basis of $\widetilde{\mathbb R}^n$ satisfy the following: For any choice of representatives $(v^i_{\varepsilon})_{\varepsilon}\;(1\leq i\leq n)$ of $v^i$ there exists some $\varepsilon_0\in I$ such that for each
$\varepsilon<\varepsilon_0$ we have
\[
\max_{i=1,\dots,n} \vert v^i_{\varepsilon}\vert>0.
\]
\item \label{freeloader6} For each representative $(v_{\varepsilon})_{\varepsilon}\in\mathcal E_M(\mathbb R^n)$ of $v$ there exists some $\varepsilon_0\in I$
such that for each $\varepsilon<\varepsilon_0$ we have $v_{\varepsilon}\neq 0$ in $\mathbb R^n$.
\item \label{freeloader7} There exists a basis of $\widetilde{\mathbb R}^n$ such that the first coefficient $v^i$ of $v$ is strictly non-zero.
\item \label{freeloader8} $v$ can be extended to a basis of $\widetilde{\mathbb R}^n$.
\item \label{freechar5} Let $v^i$ ($i=1,\dots,n$) denote the coefficients of $v$ with respect to an arbitrary basis of $\widetilde{\mathbb R}^n$. Then we have
\[
\| v\widetilde{\|}:=\left(\sum_{i=1}^n (v^i)^2\right)^{1/2}>0.
\]
\end{enumerate}
\end{theorem}
\begin{proof}
The equivalences of (\ref{freechar1}) $\Leftrightarrow$ (\ref{freechar5}) as well as (\ref{freechar4}) $\Leftrightarrow$ (\ref{freeloader6})
are evident. We start by establishing the implications (\ref{freechar1}) $\Rightarrow$ (\ref{freechar2}) $\Rightarrow$
(\ref{freechar3}) $\Rightarrow$ (\ref{freechar1}) and the equivalence (\ref{freechar4}) $\Leftrightarrow$ (\ref{freechar5}).
We end the proof by showing (\ref{freechar4}) $\Rightarrow$ (\ref{freeloader7}) $\Rightarrow$ (\ref{freeloader8}) 
$\Rightarrow$ (\ref{freechar4}).\\
If $v=0$ the equivalences are trivial. We shall therefore assume $v\neq 0$.\\
(\ref{freechar1}) $\Rightarrow$ (\ref{freechar2}): Let $(h_{ij})_{ij}$ be the coefficient matrix of $h$ with respect to some fixed basis
$\mathcal B$ of $\widetilde{\mathbb R}^n$. Then $\lambda:=\sum_{1\leq i,j\leq n}h_{ij}v^iv^j=h(v,v)>0$, in particular
$\lambda$ is invertible and $\sum_j(\sum_i \frac{h_{ij}v^i}{\lambda})v^j=1$ which shows that
$\langle\{ v^1,\dots,v^n\}\rangle=\widetilde{\mathbb R}$. Since the choice of the basis was arbitrary, (\ref{freechar2}) is shown.
\\(\ref{freechar2}) $\Rightarrow$ (\ref{freechar3}):
We assume $\langle\{ v^1,\dots,v^n\}\rangle=\widetilde{\mathbb R}$ but that there exists some $\lambda\neq 0: \lambda v=0$, that is,
$\forall\;i: 1\leq i\leq n:\,\lambda v^i=0$. Since the coefficients of $v$ span $\widetilde{\mathbb R}$,
there exist $\mu_1,\dots,\mu_n$ such that $\lambda=\sum_{i=1}^n\mu_iv^i$. It follows that
$\lambda^2=\sum_{i=1}^n\mu_i(\lambda v^i)=0$ but this is impossible, since $\widetilde{\mathbb R}$ contains no nilpotent elements.
\\(\ref{freechar3}) $\Rightarrow$ (\ref{freechar1}):
Due to Lemma \ref{schur} we may assume that we have chosen a basis such that the coefficient matrix with respect to the latter
is in diagonal form, i.\ e., $(h_{ij})_{ij}=\diag (\lambda_1,\dots,\lambda_n)$ with $\lambda_i>0\; (1\leq i\leq n)$. We have to show that $h(v,v)=\sum_{i=1}^n\lambda_i (v^i)^2>0$. Since there exists
$\varepsilon_0\in I$ such that for all representatives of $\lambda_1,\dots,\lambda_n, v^1,\dots, v^n$ we have for $\varepsilon<\varepsilon_0$ that $\gamma_{\varepsilon}:=\lambda_{1\varepsilon} (v^1_{\varepsilon})^2+\dots+\lambda_{n\varepsilon} (v^n_{\varepsilon})^2\geq 0$,
$h(v,v)\not> 0$ would imply that there exists a zero sequence $\varepsilon_k\rightarrow 0$ ($k\rightarrow \infty$) such that $\gamma_{\varepsilon_k}<\varepsilon^k$. This implies that $h(v,v)$ is a zero divisor
and it means that all summands share a simultaneous zero divisor, i.\ e., $\exists\;\mu\neq 0\,\forall\;i\in\{1,\dots,n\}:\,\mu \lambda_i (v^i)^2=0$. Since $v$ was free, this is a contradiction and we have shown that  (\ref{freechar1}) holds.\\  We proceed by establishing the equivalence
(\ref{freechar4}) $\Leftrightarrow$ (\ref{freechar5}). First, assume (\ref{freechar5}) holds, and let $(v^i_{\varepsilon})_{\varepsilon}\;(1\leq i\leq n)$ be arbitrary representatives of $v^i\;(i=1,\dots,n)$. Then
\[
\left(\sum_{i=1}^n (v^i_{\varepsilon})^2\right)_{\varepsilon}
\]
is a representative of $(\|v\widetilde{\|})^2$ as well, and since $\|v\widetilde{\|}$ is strictly positive, there exists some $m_0$ and some $\varepsilon_0\in I$ such that
\[
\forall\;\varepsilon<\varepsilon_0:\sum_{i=1}^n (v^i_{\varepsilon})^2>\varepsilon^{m_0}.
\]
This immediately implies (\ref{freechar4}). In order to see the converse direction, we proceed indirectly. Assume (\ref{freechar5}) does not hold, that is, we assume there exist representatives $(v_{\varepsilon}^i)_{\varepsilon}$ of $v^i$ for $i=1,\dots,n$ such that
for some sequence $\varepsilon_k\rightarrow 0$ ($k\rightarrow\infty$) we have for each $k>0$ that
\[
\sum_{i=1}^n (v_{\varepsilon_k}^i)^2<\varepsilon_k^k.
\]
Therefore one may even construct representatives $(\widetilde v_{\varepsilon}^i)_{\varepsilon}$ for $v^i$ ($i=1,\dots,n$)
such that for each $k>0$ and each $i\in\{1,\dots,n\}$ we have $\widetilde v_{\varepsilon_k}^i=0$. It is now evident that $(\widetilde v_{\varepsilon}^i)_{\varepsilon}$ violate condition (\ref{freechar4}) and we are done with (\ref{freechar4}) $\Leftrightarrow$ (\ref{freechar5}).
Finally we prove the chain of implications (\ref{freechar4}) $\Rightarrow$ (\ref{freeloader7}) $\Rightarrow$ (\ref{freeloader8}) $\Rightarrow$ (\ref{freechar4}). Clearly (\ref{freeloader8}) $\Rightarrow$ (\ref{freechar4}). To see (\ref{freechar4}) $\Rightarrow$ (\ref{freeloader7}), let $v_i$ ($i=1,\dots,n$) be the coefficients of $v$ with respect to some basis of $\widetilde{\mathbb R}^n$. We first observe that condition (\ref{freechar4}) implies that there exists some $m_0$
such that for suitable representatives $(v_{\varepsilon}^i)_{\varepsilon}$ of $v^i$ ($i=1,\dots,n$) we have for each $\varepsilon\in I$ $\max_{i=1,\dots,n}\vert v_{\varepsilon}^i\vert>\varepsilon^{m_0}$, i.\ e.,
\[
\forall\;\varepsilon\in I\;\exists\; i(\varepsilon)\in\{1,\dots,n\}:\vert v_{\varepsilon}^{i(\varepsilon)}\vert>\varepsilon^{m_0}.
\]
We may view $(v_{\varepsilon})_{\varepsilon}:=((v_{\varepsilon}^1,\dots,v_{\varepsilon}^n)^t)_{\varepsilon}\in\mathcal E_M(\mathbb R^n)$
as a representative of $v$ in $\mathcal E_M(\mathbb R^n)/\mathcal N(\mathbb R^n)$. Denote for each $\varepsilon\in I$ by $A_{\varepsilon}$ the representing matrix of the linear map $\mathbb R^n\rightarrow\mathbb R^n$ that merely permutes the $i(\varepsilon)$ th.\ canonical coordinate of $\mathbb R^n$ with the first one. Define $A:\widetilde{\mathbb R}^n\rightarrow\widetilde{\mathbb R}^n$ the bijective linear map with representing matrix
\[
A:=(A_{\varepsilon})_{\varepsilon}+\mathcal E_M(\mathcal M_n(\mathbb R)).
\]
What is evident now from our construction, is: The first coefficient of
\[
\widetilde v:=Av=(\mathcal A_{\varepsilon}v_{\varepsilon})_{\varepsilon}+\mathcal E_M(\mathbb R^n)
\]
is strictly nonzero and we have shown (\ref{freeloader7}). Finally we verify (\ref{freeloader7}) $\Rightarrow$ (\ref{freeloader8}). Let $\{e_i\mid 1\leq i\leq n\}$ denote the canonical basis of $\widetilde{\mathbb R}^n$. Point (\ref{freeloader7}) ensures the existence of a bijective linear map $A$ on $\widetilde{\mathbb R}^n$ such that
the first coefficient $\bar v^1$ of $\bar v=(\bar v^1,\dots,\bar v^n)^t:=Av$ is strictly non-zero; applying Proposition \ref{steinitzprop} yields another basis $\{\bar v,e_2,\dots,e_n\}$ of $\widetilde{\mathbb R}^n$. Since $A$ is bijective,
$\{v=A^{-1}\bar v,A^{-1}e_2,\dots,A^{-1}e_n\}$ is a basis of $\widetilde{\mathbb R}^n$ as well and we are done.
\end{proof}
We may add a non-trivial example of a free vector to the above characterization:
\begin{example}\rm
For $n>1$, let $\lambda_i\in\widetilde{\mathbb R}\; (1\leq i\leq n)$ have the following properties:
\begin{enumerate}
\item \label{ex1} $\lambda_i^2=\lambda_i\;\forall\; i\in\{1,\dots,n\}$
\item \label{ex2} $\lambda_i\lambda_j=0\;\forall\;i\neq j$
\item \label{ex3} $\langle\{\lambda_1,\dots,\lambda_n\}\rangle=\widetilde{\mathbb R}$
\end{enumerate}
This choice of zero divisors in $\widetilde{\mathbb R}$ is possible (idempotent elements in $\widetilde{\mathbb R}$ are thoroughly discussed in \cite{A1}, pp.\ 2221--2224). Now, let $\mathcal B=\{e_1,\dots,e_n\}$ be the canonical basis of $\widetilde{\mathbb R}^n$. Set $v:=\sum_{i=1}^n (-1)^{(i+1)(n+1)}\lambda_i e_i$ satisfies Theorem \ref{freechar} (\ref{freechar3}). Thus $v$ is free. Furthermore let $\gamma\in\Sigma_n$ be the cyclic permutation which sends $\{1,\dots,n\}$ to $\{n,1,\dots,n-1\}$. Clearly the sign of $\gamma$ is positive if and only if $n$ is odd. Define $n$ vectors $v_j\;(1\leq j\leq n)$ by $v_1:=v$, and such that $v_j$ is given by $v_j:=\sum_{k=1}^n \lambda_{\gamma^{j-1}(k)}e_k$ whenever $j>1$. Let $A$ be the matrix having the $v_j$'s as column vectors. Then
\[
\det A=\sum_{l=1}^n\lambda_l^n=\sum_{l=1}^n\lambda_l.
\]
By properties (\ref{ex1}) and (\ref{ex3}), $\det A$ is invertible. Therefore, $\mathcal B':=\{v,v_2,\dots,v_n\}$ is a basis of $\widetilde{\mathbb R}^n$, too. The reader is invited to check further equivalent properties of $v$ according to Theorem \ref{freechar}.
\end{example}
Since any symmetric bilinear form admits an orthogonal basis due to Corollary \ref{exorthbasis} we further conclude by means of Theorem
\ref{freechar}:
\begin{corollary}\label{corbilpos}
Let $b$ be a symmetric bilinear form on $\widetilde{\mathbb R}^n$. Then the following are equivalent:
\begin{enumerate}
\item \label{corbilpos1} For any free $v\in\widetilde{\mathbb R}^n$, $b(v,v)>0$.
\item $b$ is positive definite.
\end{enumerate}
\end{corollary}
For showing further algebraic properties of $\widetilde{\mathbb R}^n$ (cf.\ section \ref{semi}), also the following lemma will be crucial:
\begin{lemma}\label{posprop}
Let $h$ be a positive definite symmetric bilinear form. Then we have the following:
\begin{enumerate}
\item \label{h1} $\forall\; v\in\widetilde{\mathbb R}^n: h(v,v)\geq 0$ and $h(v,v)=0\Leftrightarrow v=0$.
\item \label{h2}Let $\mathfrak m$ be a free submodule of $\widetilde{\mathbb R}^n$. Then $h$ is a positive definite symmetric bilinear form on $\mathfrak m$.
\end{enumerate}
\end{lemma}
\begin{proof}
First, we verify (\ref{h1}): Let $v^i\;(1\leq i\leq n)$ be the coefficients of
$v$ with respect to some orthogonal basis $\mathcal B$ for $h$. Then we can write $h(v,v)=\sum_{i=1}^n\lambda_i(v^i)^2$ with $\lambda_i$ strictly positive for each $i\in\{1,\dots,n\}$. Thus $h(v,v)\geq 0$, and $h(v,v)=0$ implies $\forall\; i\in\{1\dots  n\}:v^i=0$, i.\ e., $v=0$. This finishes the proof of part (\ref{h1}).
In order to show (\ref{h2}) we first notice that by definition, any free submodule admits a basis. Let $\mathcal B_{\mathfrak m}:=\{w_1,\dots,w_k\}$ be such for $\mathfrak m$ and denote by $h_{\mathfrak m}$ the restriction of $h$ to $\mathfrak m$. Then, due to Theorem \ref{freechar} (\ref{freechar1}), we have for all $1\leq i\leq k$, $h_{\mathfrak m}(w_i,w_i)>0$. Let $A:=(h_{\mathfrak m}(w_i,w_j))_{ij}$ be the coefficient matrix of $h_{\mathfrak m}$ with respect to $\mathcal B_{\mathfrak m}$. Since $h_{\mathfrak m}$ is symmetric, so is the matrix $A$
and thus, due to Lemma \ref{schur} there is an orthogonal matrix $U\in\widetilde{\mathbb R}^{k^2}$ and there are generalized numbers $\lambda_i\;(1\leq i\leq k)$
such that $UAU^t=\diag(\lambda_1,\dots,\lambda_k)$ which implies that the (orthogonal, thus non-degenerate) $U$ maps
$\mathcal B_{\mathfrak m}$ on an orthogonal basis $\mathcal B:=\{e_1,\dots,e_k\}$ of $\mathfrak m$ with respect to $h_{\mathfrak m}$ and again by Theorem \ref{freechar} (\ref{freechar1}) we have $\lambda_i>0\;(1 \leq i\leq k)$. By Definition \ref{bilformdef}, $h_{\mathfrak m}$ is also positive definite on $\mathfrak m$ and we are done.
\end{proof}
Since any time-like or space-like vector is free, we further have as a consequence of Theorem \ref{freechar}:
\begin{proposition}\label{coraustausch}
Suppose we are given a bilinear form of Lorentzian signature on $\widetilde{\mathbb R}^n$ and let $u\in\widetilde{\mathbb R}^n\setminus \{0\}$ be time-like, null or space-like. Then $u$ can be extended to a basis of $\widetilde{\mathbb R}^n$.
\end{proposition}
In the case of a time-like vector we know a specific basis in which the first coordinate is invertible:
\begin{remark}
Suppose we are given a bilinear form $b$ of Lorentzian signature on $\widetilde{\mathbb R}^n$, let $u$ be a time-like vector.
Due to the definition of $g$ we may suppose that we have a basis so that the scalar product of $u$ takes the form
\[
g(u,u)=-\lambda_1 (u^1)^2+\lambda_2 (u^2)^2\dots+\lambda_n (u^n)^2.
\]
with $\lambda_i$ strictly positive for each $i=1,\dots,n$. Since $g(u,u)<0$, we see that the first coordinate $u^1$
of $u$ must be strictly non-zero.
\end{remark}
It is worth mentioning that an analogue of the well known criterion of positive definiteness of matrices in $\mathcal M_n(\mathbb R)$ holds in our setting:
\begin{lemma}\label{criterion}
Let $A\in\widetilde{\mathbb R}^{n^2}$ be symmetric. If the determinants of all principal subminors of $A$ (that are the submatrices $A^{(k)}:=(a_{ij})_{1\leq i,j\leq k}\;(1\leq k\leq n)$) are strictly positive, then $A$ is positive definite.
\end{lemma}
\begin{proof}
Choose a symmetric representative $(A_{\varepsilon})_{\varepsilon}$ of $A$ (cf.\  Lemma \ref{symmetry}). Clearly the assumption
$\det A^{(k)}>0\; (1\leq k\leq n)$ implies that $\exists\; \varepsilon_0\;\exists\; m\;\forall\; k: 1\leq k\leq n\; \forall\; \varepsilon<\varepsilon_0:\det A^{(k)}_{\varepsilon}\geq \varepsilon^m$, that is, for each sufficiently small $\varepsilon$, $A_{\varepsilon}$ is a positive definite symmetric
matrix due to a well known criterion in linear algebra. Furthermore $\det A^{(n)}=\det A>0$ implies $A$ is non-degenerate which finally shows
that $A$ is positive definite.
\end{proof}
Before we go on we note that type changing of tensors on $\widetilde{\mathbb R}^n$ by means of a non-degenerate symmetric bilinear form $g$ clearly is possible.
Moreover, given a (generalized) metric $g\in\mathcal G^0_2(X)$ on a manifold $X$ (cf.\ section \ref{introducerepseudoriemannereetconnexione}), lowering (resp.\ raising) indices
of generalized tensor fields on $X$ (resp.\ tensors on $\widetilde{\mathbb R}^n$) is compatible with evaluation on compactly supported generalized points (which actually yields the resp.\  object on $\widetilde{\mathbb R}^n$). This basically follows from Proposition 3.9 (\cite{KS1}) combined
with Theorem 3.1 (\cite{KS1}). As usual we write the covector associated to $\xi\in\widetilde{\mathbb R}^n$ in abstract index notation as $\xi_a:=g_{ab}\xi^b$. We call $\xi_i\;(i=1,\dots,n)$ the covariant components of $\xi$.\\
The following technical lemma is required in the sequel:
\begin{lemma}\label{uvfree}
Let $u,v\in\widetilde{\mathbb R}^n$ such that $u$ is free and $u^tv=0$. Then for each representative $(u_\varepsilon)_\varepsilon$ of $u$ there exists a representative $(v_\varepsilon)_\varepsilon$ of $v$
such that for each $\varepsilon \in I$ we have $u^t_\varepsilon v_\varepsilon=0$.
\end{lemma}
\begin{proof}
Let $(u_\varepsilon)_\varepsilon$, $(\hat v_\varepsilon)_\varepsilon$ be representatives of $u,v$ respectively. Then there exists $(n_\varepsilon)_\varepsilon\in\mathcal N$ such that
\[
(u_\varepsilon^t)_\varepsilon(\hat v_\varepsilon)_\varepsilon=(n_\varepsilon)_\varepsilon.
\]
By Theorem \ref{freechar} (\ref{freechar4}) we conclude
\[
\exists\; \varepsilon_0\;\exists\; m_0\;\forall\;\varepsilon<\varepsilon_0\;\exists\; j(\varepsilon):\;\vert u_\varepsilon^{j(\varepsilon)}\vert\geq \varepsilon^{m_0}.
\]
Therefore we may define a new representative $(v_\varepsilon)_\varepsilon$ of $v$ in the following way: For $\varepsilon\geq\varepsilon_0$ we set
$v_\varepsilon:=0$, otherwise we define
\[
v_\varepsilon:=\begin{cases}\hat v_\varepsilon^{j}, \quad j\neq j(\varepsilon)\\\hat v_{\varepsilon}^{j(\varepsilon)}-\frac{n_{\varepsilon}}{u_\varepsilon^{j(\varepsilon)}}\quad \mbox{otherwise} \end{cases}
\]
and clearly we have $u^t_\varepsilon v_\varepsilon=0$ for each $\varepsilon\in I$.
\end{proof}
The following result in the style of \cite{FL1} (Lemma 3.1.1, p.\ 74) prepares the inverse Cauchy-Schwarz inequality in our framework.
We follow the book of Friedlander which helps us to calculate the determinant of the coefficient matrix of a symmetric bilinear form,
which then turns out to be strictly positive, thus invertible. This is equivalent to non-degenerateness of the bilinear form
(cf.\ Lemma \ref{nondeg}):
\begin{proposition}\label{procs}
Let $g$ be a symmetric bilinear form of Lorentzian signature. If $u\in\widetilde{\mathbb R}^n$ is time-like, then $u^{\perp}$
is an $n$$-$$1$ dimensional submodule of $\widetilde{\mathbb R}^n$ and $g\mid_{u^{\perp}\times u^{\perp}}$ is positive definite.
\end{proposition}
\begin{proof}
Due to Proposition
\ref{coraustausch} we can choose a basis of $\widetilde{\mathbb R}^n$ such that $\Pi:=\langle\{u\}\rangle$ is spanned by the first vector, i.\ e.,
\[
\Pi=\{\xi\in\widetilde{\mathbb R}^n\vert \xi^A=0, A=2,\dots,n\}.
\]
Consequently we have
\[
\langle \xi,\xi\rangle\vert_{\Pi\times\Pi}=g_{11}(\xi^1)^2,
\]
and $g_{11}=\langle u,u\rangle<0$. If $\eta\in \Pi':=u^{\perp}$, then $\langle \xi,\eta\rangle=\xi^i\eta_i$, hence the covariant component $\eta_1$ must vanish (set $\xi:=u$, i.\ e., $\langle \xi,\eta\rangle=\langle u,\eta\rangle=\eta_1=0$). Therefore we have
\begin{equation}\label{uperp}
\langle \eta ,\theta\rangle\vert_{\Pi'\times\Pi'}=g^{AB}\eta_A\theta_B.
\end{equation}
Our first observation is that $u^{\perp}$ is a free ($n-1$ dimensional) submodule
with the basis $\xi_{(2)},\dots,\xi_{(n)}$ given in terms of the chosen coordinates above
via
\[
\xi_{(k)}^j:=g^{ij}\delta_i^k,\quad k=2,\dots,n
\]
(cf.\ (\ref{matmulti}) below, these are precisely the $n-1$ row vectors there!)
Due to the matrix multiplication
\begin{equation}\label{matmulti}
\left (\begin{array}{cccc} 1 & 0 &\dots &0 \\ g^{21}& g^{22}&\dots&g^{2n}\\ \dots &\dots&\dots&\dots\\g^{n1}& g^{n2}&\dots&g^{nn}\end{array}\right)(g_{ij})=\left(\begin{array}{cc} g_{11}& *\\ 0&\mathbb I_{n-1}\end{array} \right)
\end{equation}
evaluation of the determinants yields
\[
\det g^{AB}\det g_{ij}=g_{11}.
\]
And it follows from $\det g_{ij}<0, g_{11}<0$ that $\det g^{AB}>0$ which in particular shows that $g^{AB}$ is a non-degenerate symmetric matrix, $g\mid_{u^{\perp}\times u^{\perp}}$ therefore being a non-degenerate symmetric bilinear form on an $n-1$ dimensional free submodule. What is left to prove is positive definiteness of $g^{AB}$.
We claim that for each  $u\in v^{\perp}$, $g(v,v)\geq 0$. In conjunction with the fact that $g\mid_{u^{\perp}\times u^\perp}$ is non-degenerate, it follows that $g(v,v)>0$ for any free $v\in u^{\perp}$ (this can be seen by using a suitable basis for $u^{\perp}$ which diagonalizes $g\mid_{u^{\perp}\times u^{\perp}}$, cf.\ Corollary \ref{corbilpos}) and we are done.

To show the subclaim we have to undergo an $\varepsilon$-wise argument. Let $(u_{\varepsilon})_{\varepsilon}\in\mathcal E_M(\mathbb R^n)$ be a representative of $u$ and let
$((g^{\varepsilon}_{ij})_{ij})_{\varepsilon}\in\mathcal E_M(\mathcal M_n(\mathbb R))$ be a symmetric representatives of $(g_{ij})_{ij}$, where $(g_{ij})_{ij}$ is the coefficient matrix of $g$ with respect to the canonical basis of $\widetilde{\mathbb R}^n$. For each $\varepsilon$ we denote by $g_{\varepsilon}$ the symmetric bilinear form
induced by $(g^{\varepsilon}_{ij})_{ij}$, that is, the latter shall be the coefficient matrix of $g_{\varepsilon}$ with respect to the canonical basis of $\mathbb R^n$. First we show that
\begin{equation}\label{identitynormalspaces}
u^{\perp}=\{(v_{\varepsilon})_{\varepsilon}\in\mathcal E_M(\mathbb R^n):\; \forall\; \varepsilon>0: v_{\varepsilon}\in u_{\varepsilon}^{\perp}\}+\mathcal N(\mathbb R^n),
\end{equation}
Since the inclusion relation $\supseteq$ is clear, we only need to show that $\subseteq$ holds. To this end,
pick $v\in u^{\perp}$. Then $g(u,v)=g_{ij}u^iv^j=0$ and the latter implies that for each representative $(\hat v_{\varepsilon})_{\varepsilon}$ of $v$ there exists $(n_{\varepsilon})_{\varepsilon}\in\mathcal N$
such that
\[
(g_{ij}^{\varepsilon}u_{\varepsilon}^i\hat v_{\varepsilon}^j)_{\varepsilon}=(n_{\varepsilon})_{\varepsilon}.
\]
We may interpret $(g_{ij}^{\varepsilon}u_{\varepsilon}^i) (j=1,\dots,n)$ as the representatives of the coefficients of a vector $w$ with coordinates  $w_j:=g_{ij}u^i$, and $w$ is free, since $u$ is free and $g$ is non-degenerate. Therefore we may employ Lemma \ref{uvfree} which yields a representative $(v_{\varepsilon}^j)_{\varepsilon}$ of $v$
such that
\[
(g_{ij}^{\varepsilon}u_{\varepsilon}^i
 v_{\varepsilon}^j)_{\varepsilon}=0.
\]
This precisely means that there exists a representative $(v_{\varepsilon})_\varepsilon$ of $v$ such
that for each $\varepsilon$ we have $v_\varepsilon\in u_\varepsilon^\perp$. We have thus finished the proof of identity (\ref{identitynormalspaces}).

To finish the proof of the claim, that is $g(v,v)\geq0$, we pick a representative $(v_{\varepsilon})_{\varepsilon}$ of $v$ and an $\varepsilon_0\in I$ such that for each
$\varepsilon<\varepsilon_0$ we have
\begin{enumerate}
\item each $g_{\varepsilon}$ is of Lorentzian signature
\item $u_{\varepsilon}$ is time-like
\item $v_{\varepsilon}\in u_{\varepsilon}^\perp$.
\end{enumerate}
Note that this choice is possible due to (\ref{identitynormalspaces}). Further, by the resp.\ classic result of Lorentz geometry (cf.\ \cite{FL1}, Lemma 3.\ 1.\ 1) we have
$g_{\varepsilon}(v_{\varepsilon},v_\varepsilon)\geq 0$ unless $v_{\varepsilon}=0$. Since
$(g_{ij}^{\varepsilon}v_\varepsilon^i v_\varepsilon^j)_\varepsilon$ is a representative of
$g(v,v)$ we have achieved the subclaim.
\end{proof}
\begin{corollary}\label{dirsum1}
Let $u\in\widetilde{\mathbb R}^n$ be time-like. Then $u^{\perp}:=\{v\in \widetilde{\mathbb R}^n:\langle u,v\rangle=0\}$ is a submodule of $\widetilde{\mathbb R}^n$ and  $\widetilde{\mathbb R}^n=\langle \{u\}\rangle \oplus u^{\perp}$.
\end{corollary}
\begin{proof}
The first statement is obvious. For $v\in\widetilde{\mathbb R}^n$, define the orthogonal projection of $v$ onto $\langle\{u\}\rangle$ as $P_u(v):=\frac{\langle u, v\rangle}{\langle u,u\rangle}u$. Then one sees that $v=P_u(v)+(v-P_u(v))\in \langle \{u\}\rangle+u^{\perp}$.
Finally, assume $\widetilde{\mathbb R}^n\neq \langle \{u\}\rangle \oplus u^{\perp}$, i.\ e., $\exists\; \xi\neq 0, \xi\in \langle \{u\}\rangle \cap u^{\perp} $. It follows $\langle \xi,\xi\rangle\leq 0$ and due to the preceding proposition $\xi\in u^{\perp}$ implies $\langle \xi,\xi\rangle \geq 0$. Since we have a partial ordering $\leq $, this is impossible unless $\langle \xi,\xi\rangle=0$. However by Lemma \ref{posprop} (\ref{h1}) we have $\xi=0$. This contradicts our assumption and proves that $\widetilde{\mathbb R}^n$ is the direct sum of $u$ and its orthogonal complement.
\end{proof}
The following statement on the Cauchy--Schwarz inequality is a crucial result in generalized Lorentz Geometry. It slightly differs from the classical result as is shown in Example \ref{csex}. However it seems to coincide with the classical inequality in physically relevant cases, since
algebraic complications which mainly arise from the existence of zero divisor in our scalar ring
of generalized numbers, presumably are not inherent in the latter. Our proof follows the lines of the proof of the analogous classic statement in O'Neill's book (\cite{ON}, chapter 5, Proposition 30, pp.\ 144):
\begin{theorem}(Inverse Cauchy--Schwarz inequality)\label{cs}
Let $u,\;v \in \widetilde{\mathbb R}^n$ be time-like vectors. Then
\begin{enumerate}
\item \label{cs1} $\langle u,v\rangle^2\geq \langle u,u\rangle \langle v,v\rangle$, and
\item \label{cs2} equality in (\ref{cs1}) holds if $u,v$ are linearly dependent over $\widetilde{\mathbb R}^*$, the units
in $\widetilde{\mathbb R}$.
\item \label{cs3} If $u,v$ are linearly independent, then $\langle u,v\rangle^2>\langle u,u\rangle \langle v,v\rangle$.
\end{enumerate}
\end{theorem}
\begin{proof}
In what follows, we keep the notation of the preceding corollary. Due to Corollary \ref{dirsum1}, we may decompose $u$ in a unique way
$v=a u+w$ with $a\in\widetilde{\mathbb R},\, w\in u^{\perp}$. Since $u$ is time-like,
\[
\langle v,v\rangle=a^2 \langle u,u\rangle+\langle w,w\rangle<0.
\]
Then
\begin{equation}\label{eqcs}
\langle u,v\rangle^2=a^2\\\langle u,u\rangle^2=(\langle v,v\rangle-\langle w,w\rangle)\langle u,u\rangle\geq \langle u,u\rangle \langle v,v\rangle
\end{equation}
since $\langle w,w\rangle\geq 0$ and this proves (\ref{cs1}). \\In order to prove (\ref{cs2}), assume $u,v$ are linearly dependent over $\widetilde{\mathbb R}^*$, that is, there exist $\lambda,\,\mu$, both units in $\widetilde{\mathbb R}$ such that
$\lambda u+\mu v=0$. Then $u=-\frac{\mu}{\lambda} v$ and equality in (\ref{cs2}) follows.\\ Proof of (\ref{cs3}): Assume now, that
$u,v$ are linearly independent. We show that this implies that $w$ is free.
For the sake of simplicity we assume without loss of generality that $\langle u,u\rangle=\langle v,v\rangle=-1$ and we choose
a basis $\mathcal B=\{e_1,\dots,e_n\}$ with $e_1=u$ due to Proposition \ref{coraustausch}.
Then with respect to the new basis we can write $u=(1,0,\dots,0)^t$, $v=(v^1,\dots,v^n)^t$, $w=v-P_u(v)=(v^1-(-g(v,e_1)), v^2,\dots,v^n)^t=(0,w^2,\dots,w^n)^t$.
Assume $\exists\; \lambda\neq0: \lambda w=0$, then
\[
(\lambda v^1)u+\lambda v=\lambda v^1 e_1-\lambda g(v,e_1) e_1=\lambda v^1 e_1-\lambda v^1 e_1=0
\]
which implies that $u,v$ are linearly dependent. This contradicts the assumption in (\ref{cs3}). Thus $w$ indeed is free.
Applying Theorem \ref{freechar} yields $\langle w,w\rangle>0$. A glance at
(\ref{eqcs}) shows that the proof of (\ref{cs3}) is finished.
\end{proof}
The following example indicates what happens when in \ref{cs} (\ref{cs2}) linear dependence over the units in $\widetilde{\mathbb R}$ is replaced by
linear dependence over $\widetilde{\mathbb R}$:
\begin{example}\label{csex}
Let $\lambda\in\widetilde{\mathbb R}$ be an idempotent zero divisor, and write $\alpha:=[(\varepsilon)_{\varepsilon}]$. Let $\eta=\diag(-1,1\dots,1)$ be the Minkowski metric.
Define $u=(1,0,\dots,0)^t,v=(1,\lambda\alpha,0,\dots,0)^t$. Clearly $\langle u,u\rangle=-1,\langle v,v\rangle=-1+\lambda^2\alpha^2<0$
But
\[
\langle u,v\rangle^2=1\neq \langle u,u\rangle\langle v,v\rangle=-(-1+\lambda^2\alpha^2)=1-\lambda^2\alpha^2.
\]
However, also the strict relation fails, i.\ e., $\langle u,v\rangle^2 \not> \langle u,u\rangle\langle v,v\rangle$, since $\lambda$
is a zero divisor.
\end{example}
\subsection{Applications}\label{energygeneralizedsection} 
In this subsection we establish a generalized dominant energy condition as an application of the
Cauchy-Schwarz inequality. This generalizes known results
in Relativity by Hawking and Ellis (\cite{HE}) in the context of the special algebra.
Throughout this subsection $g$ denotes a symmetric bilinear form of Lorentz signature
on $\widetilde{\mathbb R}^n$, and for $u,v\in\widetilde{\mathbb
R}^n$ we write $\langle u,v\rangle:=g(u,v)$. We introduce the notion
of a (generalized) Lorentz transformation:
\begin{definition}
We call a linear map $L:\widetilde{\mathbb R}^n\rightarrow\widetilde{\mathbb R}^n$ a Lorentz transformation, if it preserves
the metric, that is
\[
\forall \xi \in \widetilde{\mathbb R}^n:\;\langle L\xi,L\eta\rangle=\langle \xi,\eta\rangle
\]
or equivalently,
\[
L^{\mu}_{\lambda}L^{\nu}_{\rho}g_{\mu\nu}=g_{\lambda\rho}.
\]
\end{definition}
The following statement is straightforward (cf.\ \cite{RB}):
\begin{lemma}\label{lorentz}
Let $\xi, \eta\in \widetilde{\mathbb R}^n$ be time-like unit vectors with the same time-orientation. Then
\[
L^{\mu}_{\lambda}:=\delta^{\mu}_{\lambda}-2\eta^{\mu}\xi_{\lambda}+\frac{(\xi^{\mu}+\eta^{\mu})(\xi_{\lambda}+\eta_{\lambda})}{1-\langle \xi,\eta\rangle}
\]
is a Lorentz transformation with the property $L\xi=\eta$.
\end{lemma}
The following proposition is a crucial ingredient in the subsequent proof of the (generalized) dominant energy condition for certain energy tensors of this section:
\begin{proposition}\label{metrconstr}
Let $u,v \in \widetilde{\mathbb R}^n$ be time-like vectors such that $\langle u,v\rangle<0$. Then
\[
h_{\mu\nu}:=u_{(\mu}v_{\nu)}-\frac{1}{2}\langle u,v\rangle g_{\mu\nu}
\]
is a positive definite symmetric bilinear form on $\widetilde{\mathbb R}^n$.
\end{proposition}
\begin{proof}
Symmetry and bilinearity of $h$ are clear. What would be left is to show that the coefficient matrix of $h$ with respect to an arbitrary basis is invertible.
However, determining the determinant of $h$ is nontrivial. So we proceed by showing that for any free $w\in\widetilde{\mathbb R}^n$,
$h(w,w)$ is strictly positive (thus also deriving the classic statement). We may assume $\langle u ,u\rangle=\langle v,v\rangle=-1$; this can be achieved by scaling $u,v$ (note that this is due to the fact that for a time-like (resp.\ space-like) vector $u$, $\langle u, u\rangle$ is strictly non-zero, thus invertible in $\widetilde{\mathbb R}$). We may assume we have chosen an orthogonal basis $\mathcal B=\{e_1,\dots,e_n\}$ of $\widetilde{\mathbb R}^n$ with respect to $g$, i.\ e., $g(e_i,e_j)=\varepsilon_{ij}\lambda_i$, where $\lambda_1\leq \dots\leq \lambda_n$  are the eigenvalues of $(g(e_i,e_j))_{ij}$.
Due to Lemma \ref{lorentz} we can treat $u,v$ by means of generalized Lorentz transformations such that both vectors appear in the form $u=(\frac{1}{\lambda_1},0,0,0)$, $v=\gamma(v)(\frac{1}{\lambda_1},\frac{V}{\lambda_2},0,0)$, where $\gamma(v)=\sqrt{-g(v,v)}=\sqrt{1-V^2} >0$ (therefore $\vert V\vert <1$). Let $w=(w^1,w^2,w^3,w^4)\in \widetilde{\mathbb R}^n$ be free (in particular $w\neq 0$). Then
\begin{equation}\label{beig1}
h(w,w):=h_{ab}w^aw^b= \langle u,w\rangle  \langle v,w\rangle-\frac{1}{2}\langle w,w\rangle \langle u,v\rangle.
\end{equation}
Obviously, $\langle u,w \rangle=-w^1,\langle v,w\rangle=\gamma(v)(-w^1+Vw^2),\langle u,v\rangle=-\gamma(v)$. Thus
\begin{multline}\nonumber
h(w,w)=\gamma(v)(-w^1)(-w^1+Vw^2)+\frac{\gamma(v)}{2}(-(w^1)^2+(w^2)^2+(w^3)^2+(w^4)^2)=\\=
-\gamma(v) V w^1w^2+\frac{1}{2}\gamma(v) (+(w^1)^2+(w^2)^2+(w^3)^2+(w^4)^2).
\end{multline}
If $Vw^1w^2\leq 0$, we are done. If not, replace $V$ by $\vert V\vert$ ($-V\geq-\vert V \vert$) and rewrite the last formula in the following form :
\begin{equation}\label{estw}
h(w,w)\geq\frac{\gamma(v)}{2}\left ( (\vert V \vert(w^1-w^2)^2 +(1-\vert V \vert)(w^1)^2+(1-\vert V \vert)(w^2)^2+(w^3)^2+(w^4)^2\right).
\end{equation}
Clearly for the first term on the right side of (\ref{estw}) we have $\vert V\vert(w^1-w^2)^2\geq 0$. From $v$ is time-like we further deduce $1-\vert V \vert=\frac{1-V^2}{1+\vert V \vert}>0$. Since $w$ is free we may apply Theorem \ref{freechar}, which yields $(1-\vert V\vert)(w^1)^2+(1-\vert V\vert)(w^2)^2+(w^3)^2+(w^4)^2>0$
and thus $h(w,w)>0$ due to equation (\ref{estw} and we are done.
\end{proof}
Finally we are prepared to show a dominant energy condition in the style of Hawking and Ellis (\cite{HE}, pp.\ 91--93) for a generalized energy tensor. In what follows, we use abstract index notation.
\begin{theorem}\label{dec}
For $\theta\in\widetilde{\mathbb R}^n$ the energy tensor $E^{ab}(\theta):=(g^{ac}g^{bd}-\frac{1}{2}g^{ab}g^{cd})\theta_c\theta_d$ has the following properties
\begin{enumerate}
\item \label{energy1} If $\xi,\eta\in\widetilde{\mathbb R}^n$ are time-like vectors with the same orientation, then we have for any free
$\theta$, $E^{ab}(\theta)\xi_a\eta_b>0$.
\item \label{energy2} Suppose $\langle \theta,\theta\rangle$ is invertible in $\widetilde{\mathbb R}$. If $\xi\in\widetilde{\mathbb R}^n$ is time-like, then $\eta^b:=E^{ab}(\theta)\xi_a$ is time-like and $\eta^a\xi_a>0$, i.\ e., $\eta$ is past-oriented with respect to $\xi$. Conversely, if $\langle \theta,\theta\rangle$ is a zero divisor, then $\eta$ fails to be time-like.
\end{enumerate}
\end{theorem}
\begin{proof}
(\ref{energy1}): Define a symmetric bilinear form  $h^{ab}:=(g^{(ac}g^{b)d}-\frac{1}{2}g^{ab}g^{cd})\xi_c\eta_d$. Due to our assumptions on
$\xi$ and $\eta$, Proposition \ref{metrconstr} yields that $h^{ab}$ is a positive definite symmetric bilinear form. By Theorem \ref{freechar} we conclude that for any free $\theta\in\widetilde{\mathbb R}^n$, $h_{ab}\theta^a\theta^b>0$. It is not hard to check that $E^{ab}(\theta)\xi_a\eta_b=h^{ab}\theta_a\theta_b$ and therefore we have proved (\ref{energy1}).\\ (\ref{energy2}): To start with, assume $\eta$ is time-like. Then $g(\xi,\eta)=g_{ab}\xi^a\eta^b=g_{ab}\xi^aE(\theta)^{ac}\xi_c=E^{ab}(\theta)\xi_a\xi_b$.
That this expression is strictly greater than zero follows from (\ref{energy1}), i.\ e., $E^{ab}(\theta)\xi_a$ is past-directed with respect to $\xi$
whenever $\langle\theta,\theta\rangle$ is invertible, since the latter implies $\theta$ is free. It remains to prove that $\langle \eta,\eta\rangle<0$. A straightforward calculation yields
\[
\langle \eta,\eta\rangle=\langle E(\theta)\xi,E(\theta)\xi\rangle=\frac{1}{4}\langle \theta,\theta\rangle^2\langle \xi,\xi\rangle.
\]
Since $\langle \theta,\theta\rangle$ is invertible and $\xi$ is time-like, we conclude that $\eta$ is time-like as well. Conversely, if $\langle \theta,\theta\rangle$ is a zero-divisor, also $\langle E(\theta)\xi,E(\theta)\xi\rangle$ clearly is one. Therefore, $\eta=E(\theta)\xi$ cannot be time-like, and we are done.
\end{proof}
A remark on this statement is in order. A comparison with (\cite{HE}, pp.\ 91--93) shows, that
our \lq\lq dominant energy condition\rq\rq\ on $T^{ab}$ is stronger, since the vectors $\xi,\eta$ in (\ref{energy1}) need not coincide. Furthermore,
if in (\ref{energy2}) the condition \lq\lq $\langle \theta,\theta\rangle$ is invertible\rq\rq\ was dropped, then (as in the classical (\lq\lq smooth\rq\rq\ ) theory) we could conclude that $\eta$ was not space-like, however, unlike in the smooth theory, this does not imply $\eta$ to be time-like or null (cf.\ the short note after Definition \ref{causaldef1}).
\section[Point values and generalized causality]{Generalized point value characterizations of generalized pseudo-Riemannian metrics and of causality of generalized vector fields}\label{sec6}
The first goal of this section is to characterize generalized pseudo-Riemannian metrics through evaluation on generalized points.
Then we describe causality of generalized vector fields on $X$ by means of
causality in $\widetilde{\mathbb R}^n$. The importance of the latter comes from the fact that generalized functions are not uniquely determined
by evaluation on standard points (cf.\ the discussion in subsection \ref{uniqueness}). We start by establishing a point-value characterization of generalized pseudo-Riemannian metrics with respect to their index:
\begin{theorem}\label{charindexg}
Let $g\in \mathcal G^0_2(X)$ satisfy one (hence all) of the equivalent statements of Theorem \ref{chartens02}, $j\in\mathbb N_0$. The following are equivalent:
\begin{enumerate}
\item \label{charindexg1} $g$ has (constant) index $j$.
\item \label{charindexg2} For each chart $(V_{\alpha},\psi_{\alpha})$ and each $\widetilde x \in (\psi_{\alpha}(V_{\alpha}))_c^{\sim}$, $g_{\alpha}(\widetilde x)$ is a symmetric bilinear form on $\widetilde{\mathbb R}^n$ with index $j$.
\end{enumerate}
\end{theorem}
\begin{proof}
(\ref{charindexg1})$\Rightarrow$(\ref{charindexg2}): Let $\widetilde x\in\psi_{\alpha}(V_{\alpha})_c^{\sim}$ be supported in $K\subset\subset \psi_{\alpha}(V_{\alpha})$ and choose a representative $(g_{\varepsilon})_{\varepsilon}$ of $g$ as in Theorem \ref{chartens02} (\ref{chartens023}) and Definition \ref{defpseud}. According to Theorem \ref{chartens02} (\ref{chartens021}), $g_{\alpha}(\widetilde x):\widetilde{\mathbb R}^n\times\widetilde{\mathbb R}^n\rightarrow\widetilde{\mathbb R}$ is symmetric and non-degenerate. So it merely remains to prove that
the index of $g_\alpha(\widetilde x)$ coincides with the index of $g$. Since $\widetilde x$ is compactly supported, we may shrink $V_{\alpha}$
to $U_\alpha$ such that the latter is an open relatively compact subset of $X$ and $\widetilde x\in \psi_\alpha(U_\alpha)$. By Definition \ref{defpseud}
there exists a symmetric representative $(g_\varepsilon)_\varepsilon$ of $g$ on $U_\alpha$ and an $\varepsilon_0$ such that for all
$\varepsilon<\varepsilon_0$, $g_\varepsilon$ is a pseudo-Riemannian metric on $U_\alpha$ with constant index $\nu$. Let $(\widetilde x_\varepsilon)_\varepsilon$
be a representative of $\widetilde x$ lying in $U_\alpha$ for each $\varepsilon<\varepsilon_0$. Let $g_{\alpha,\,ij}^\varepsilon$
 be the coordinate expression of $g_\varepsilon$ with respect to the chart $(U_\alpha,\psi_\alpha)$. Then for each $\varepsilon<\varepsilon_0$, $g_{\alpha,\,ij}^\varepsilon(\widetilde x_\varepsilon)$ has precisely $\nu$ negative and $n-\nu$ positive eigenvalues,
 therefore due to Definition \ref{indexmatrixdef}, the class $g_{ij}:=[(g_{\alpha,\,ij}^\varepsilon(\widetilde x_\varepsilon))_\varepsilon]\in\mathcal M_n(\widetilde {\mathbb R})$
 has index $\nu$. By Definition \ref{bilformdef} it follows that the respective bilinear form $g_{\alpha}(\widetilde x)$ induced by $(g_{ij})_{ij}$ with respect to the canonical basis of $\widetilde{\mathbb R}$ has index $\nu$ and we are done.\\ To show the converse direction, one may proceed
 by an indirect proof. Assume the contrary to (\ref{charindexg1}), that is, $g$ has non-constant index $\nu$. In view of Definition \ref{defpseud}
 there exists an open, relatively compact chart $(V_\alpha, \psi_\alpha)$, a symmetric representative $(g_\varepsilon)_\varepsilon$ of $g$ on $V_\alpha$
and a zero sequence $\varepsilon_k$ in $I$ such that the sequence $(\nu_k)_k$ of indices $\nu_k$ of $g_{\varepsilon_k}\mid_{V_\alpha}$
has at least two accumulation points, say $\alpha\neq\beta$. Let $(x_\varepsilon)_\varepsilon$ lie in $\psi_\alpha(V_\alpha)$ for each $\varepsilon$.
Therefore the number of negative eigenvalues of $(g_{ij})_{ij}:=(g_{\alpha,ij}^\varepsilon(x_\varepsilon))_{ij}$ is not constant for sufficiently small $\varepsilon$, and
therefore for $\widetilde x:=[(x_\varepsilon)_\varepsilon]$, the respective bilinear form $g_{\alpha}(\widetilde x)$ induced by $(g_{ij})_{ij}$ with respect to the canonical basis of $\widetilde{\mathbb R}$ has no index and we are done.
\end{proof}
Theorem \ref{downhilliseasier} provides the appropriate machinery to characterize causality of generalized vector fields:
\begin{theorem}\label{characterizationcausality}
Let ${\xi}\in\mathcal G^1_0(X)$, $ g\in \mathcal G^0_2(X)$ be a Lorentzian metric. The following are equivalent:
\begin{enumerate}
\item \label{causalitychar1} For each chart $(V_{\alpha},\psi_{\alpha})$ and each $\widetilde x \in (\psi_{\alpha}(V_{\alpha}))_c^{\sim}$, ${\xi}_{\alpha}(\widetilde x)\in\widetilde{\mathbb R}^n$ is time-like (resp.\ space-like, resp.\ null)
with respect to $ g_{\alpha}(\widetilde x)$ (a symmetric bilinear form on $\widetilde{\mathbb R}^n$ of Lorentz signature).
\item \label{causalitychar2} $ g({\xi},\xi)< 0$  (resp.\ $>0$, resp.\ $=0$)
in $\mathcal G(X)$.
\end{enumerate}
\end{theorem}
\begin{proof}
(\ref{causalitychar2})$\Leftrightarrow$ $\forall\;\widetilde x\in X_c^{\sim}: g(\xi,\xi)(\widetilde x)<0$ (due to Theorem \ref{downhilliseasier}, (\ref{invf1extra})) $\Leftrightarrow$ for each chart $(V_{\alpha},\psi_{\alpha})$ and for all $\widetilde x_c\in \psi_{\alpha}(V_{\alpha})_c^{\sim}$ we have $g_{\alpha}(\widetilde x)(\xi_{\alpha}(\widetilde x),\xi_{\alpha}(\widetilde x))<0$ in $\widetilde{\mathbb R}$ $\Leftrightarrow$ (\ref{causalitychar1}).
\end{proof}
The preceding theorem gives rise to the following definition:
\begin{definition}\label{defcausalityglobal}
A generalized vector field $ \xi\in \mathcal G^1_0(X)$ is called time-like (resp.\ space-like, resp.\ null) if
it satisfies one of the respective equivalent statements of Theorem \ref{characterizationcausality}. Moreover, two time-like vector fields $\xi,\eta$  are said to have the same time orientation, if
$\langle \xi,\eta\rangle<0$. Due to the above, this notion is consistent with the point-wise one given in
\ref{causaldef1}.
\end{definition}
We conclude this section by harvesting constructions of generalized pseudo-\\Riemannian metrics by means of
point-wise results of the preceding section in conjunction with the point-wise characterizations of the global objects of this paper:
\begin{theorem}\label{genRiemannmetricconstrglobal}
Let $g$ be a generalized Lorentzian metric and let $\xi,\eta\in \mathcal G ^1_0(X)$ be time-like vector fields with the same time orientation. Then
\[
h_{ab}:=\xi_{(a}\eta_{b)}-\frac{1}{2}\langle \xi,\eta\rangle g_{ab}
\]
is a generalized Riemannian metric.
\end{theorem}
\begin{proof}
Use Proposition \ref{metrconstr} together with Theorem \ref{characterizationcausality} and Theorem \ref{charindexg}.
\end{proof}
A final remark on this section is in order. We based our initial considerations in this paper on Theorem \ref{chartens02}, a characerization of generalized pseudo-Riemannian metrics. Point (\ref{chartens021}) motivated us to study bilinear forms $b$ on $\widetilde{\mathbb R}^n$ in section \ref{sec5} by introducing the index of $b$. We described successfully free vectors in $\widetilde{\mathbb R}^n$, established elementary (and expected) facts in generalized Lorentz geometry, for instance the Cauchy-Schwarz inequality. It was then quite natural to return to the global objects of Theorem \ref{chartens02} we had started with and to characterize them in terms of the machinery we had developed in previous sections. This section shows that the chosen notion in linear algebra on $\widetilde{\mathbb R}^n$ matchs perfectly this setting.
\section*{Appendix. Further algebraic properties of finite dimensional modules over the ring of generalized
numbers} This section is devoted to a discussion of direct summands
of submodules inside $\widetilde{\mathbb R}^n$. The question first
involves free submodules of arbitrary dimension. However,
we establish a generalization of Theorem \ref{freechar}
(\ref{freeloader8}) not only with respect to the dimension of the
submodule; the direct summand we construct is also an orthogonal
complement with respect to a given positive definite symmetric
bilinear form. Having established this in \ref{semi}, we
subsequently show that $\widetilde{\mathbb R}^n$ is not semisimple,
i.\ e., non-free submodules in our module do not admit direct
summands.
\subsection{Direct summands of free submodules}\label{semi}
The existence of positive bilinear forms on $\widetilde{\mathbb R}^n$ ensures the existence of direct summands
of free submodules of $\widetilde{\mathbb R}^n$:
\begin{theorem}
Any free submodule $\mathfrak m$ of $\widetilde{\mathbb R}^n$ has a direct summand.
\end{theorem}
\begin{proof}
Denote by $\mathfrak m$ the free submodule in question with $\dim \mathfrak m=k$, let $h$ be a positive definite symmetric bilinear form on $\mathfrak m$ and $h_{\mathfrak m}$ its restriction to $\mathfrak m$. Now, due to Lemma \ref{posprop} (\ref{h2}), $h_{\mathfrak m}$ is a positive definite symmetric bilinear form. In particular, there exists an orthogonal basis $\mathcal B_{\mathfrak m}:=\{e_1,\dots,e_k\}$ of $\mathfrak m$ with respect to $h_{\mathfrak m}$. We further may assume that the latter one is orthonormal. Denote by $P_{\mathfrak m}$ the orthogonal projection on $\mathfrak m$ which due to the orthogonality of $\mathcal B_{\mathfrak m}$ may be written in the form
\[
P_{\mathfrak m}:\;\widetilde{\mathbb R}^n\rightarrow \mathfrak m,\; v\mapsto \sum_{i=1}^k\langle v,e_i\rangle e_i.
\]
Finally, we show ${\mathfrak  m}^{\perp}=\ker P_{\mathfrak m}$:
\begin{eqnarray}\nonumber
{\mathfrak m}^{\perp}&=&\{v\in\widetilde{\mathbb R}^n\mid \forall\; u\in \mathfrak m: h(v,u)=0\}=\\\nonumber
&=&\{v\in\widetilde{\mathbb R}^n\mid \forall\; i=1,\dots,k: h(v,e_i)=0\}=\\\nonumber
&=&\{v\in \widetilde{\mathbb R}^n\mid P_{\mathfrak m}(v)=0\}=\ker P_{\mathfrak m}.
\end{eqnarray}
Where both of the last equalities are due to the definition of $P_{\mathfrak m}$ and the fact that $B_{\mathfrak m}$ is a basis of $\mathfrak m$.
As always in modules, ${\mathfrak m}^{\perp}=\ker P_{\mathfrak m}\Leftrightarrow {\mathfrak m}^{\perp}$ is a direct summand and we are done.
An alternative end of this proof is provided by Lemma \ref{posprop}: Since we have $\mathfrak m+\mathfrak m^{\perp}=\widetilde{\mathbb R}^n$,
we only need to show that this sum is a direct one. But Lemma \ref{posprop} (\ref{h1}) shows that $0 \neq u \in \mathfrak m\cap {\mathfrak m}^{\perp}$ is absurd, since $h$ is positive definite.
\end{proof}
We thus have also shown (cf.\ Theorem \ref{freechar}):
\begin{corollary}\label{orthdecriem}
Let $w\in\widetilde{\mathbb R}^n$ be free and let $h$ be a positive definite symmetric bilinear form. Then $\widetilde{\mathbb R}^n=\langle \{w\}\rangle \oplus w^{\perp}$.
\end{corollary}
We therefore have added a further equivalent property to Theorem \ref{freechar}.
\subsection{$\widetilde{\mathbb R}^n$ is not semisimple}\label{secsemi}
In this section we show that $\widetilde{\mathbb R}^n$ is not semisimple. Recall that a module $B$ over a ring $R$ is
called simple, if $RA\neq \{0\}$ and if $A$ contains no non-trivial strict submodules. For the convenience of the reader, we recall the following fact on modules (e.\ g., see \cite{Hungerford}, p.\ 417):
\begin{theorem}\label{charsemisimple}
The following conditions on a nonzero module $A$ over a ring $R$ are equivalent:
\begin{enumerate}
\item \label{charsemisimple1} $A$ is the sum of a family of simple submodules.
\item \label{charsemisimple2} $A$ is the direct sum of a family of simple submodules.
\item For every nonzero element a of $A$, $Ra\neq 0$; and every submodule $B$ of $A$ is a direct summand (that is, $A=B\oplus C$ for some submodule $C$.
\end{enumerate}
\end{theorem}
Such a module is called semisimple. However, property (\ref{charsemisimple1}) is violated in $\widetilde{\mathbb R}^n$ $(n\geq 1)$:
\begin{proposition}
Every submodule $A\neq \{0\}$ in $\widetilde{\mathbb R}^n$ contains a strict submodule.
\end{proposition}
\begin{proof}
Let $u\in A$, $u\neq 0$. We may write $u$ in terms of the canonical basis $e_i\;(i=1,\dots,n)$, $u=\sum_{i=1}^n \lambda_ie_i$ and without loss of generality we may assume $\lambda_1\neq 0$. Denote a representative of $\lambda_1$ by $(\lambda_1^{\varepsilon})_{\varepsilon}$.
$\lambda_1\neq 0$ in particular ensures the existence of a zero sequence  $\varepsilon_k \searrow 0$ in $I$ and an $m>0$ such that
for all $k\geq 1$, $\vert \lambda_1^{\varepsilon_k}\vert\geq \varepsilon_k^m$. Define $D:=\{\varepsilon_k\mid k\geq 1\}\subset I$, let
$\chi_D\in\widetilde{\mathbb R}$ be the characteristic function on $D$. Clearly, $\chi_D u\in A$, furthermore, if the submodule generated by $\chi_D u$ is not a strict submodule of $A$, one may replace $D$ by $\bar D:=\{\varepsilon_{2k}\mid k\geq 1\}$ to achieve one in the same way, which however is a strict submodule of $A$ and we are done.
\end{proof}
The preceding proposition in conjunction with Theorem \ref{charsemisimple} gives rise to the following conclusion:
\begin{corollary}
$\widetilde{\mathbb R}^n$ is not semisimple.
\end{corollary}
Our discussion on algebraic properties of the finite dimensional module $\widetilde{\mathbb R}^n$ in this paper lets us draw the following important conclusion. Though there are obvious differences to linear algebra in $\widetilde{\mathbb R}^n$, important facts still hold in the generalized setting. In particular, what is said in this section indicates that properties for linear subspaces of $\mathbb R^n$, have appropriate counterparts for {\it free} submodules of $\widetilde{\mathbb R}^n$. 
\section*{Acknowledgement}
I want to express my gratitude to my supervisors Michael Kunzinger and Roland Steinbauer
for the great research environment they have offered me inside the
DIANA research group at the University of Vienna during the last 3 years.
I am further indebted to Professor Robert Beig for discussions on
section \ref{sec5}, particularly on generalized Energy tensors.
\bibliographystyle{siam}

\begin{thebibliography}{10}

\bibitem{A1}
{\sc J.~Aragona and S.~O. Juriaans}, {\em Some structural properties of the
  topological ring of {C}olombeau's generalized numbers}, Comm. Algebra, 29
  (2001), pp.~2201--2230.

\bibitem{RB}
{\sc R.~Beig}, {\em Lecture notes on special and general relativity,
  unpublished}, University of Vienna, Physics Institute,  (2004).

\bibitem{Clarke}
{\sc C.~J.~S. Clarke}, {\em Generalized hyperbolicity in singular spacetimes},
  Class. Quantum Grav., 15 (1998), pp.~975--984.

\bibitem{CVW}
{\sc C.~J.~S. Clarke, J.~A. Vickers, and J.~P. Wilson}, {\em Generalized
  functions and distributional curvature of cosmic strings}, Class. Quantum
  Grav., 13 (1996), pp.~2485--2498.

\bibitem{Colombeau}
{\sc J.-F. Colombeau}, {\em New generalized functions and multiplication of
  distributions}, vol.~84 of North-Holland Mathematics Studies, North-Holland
  Publishing Co., Amsterdam, 1984.
\newblock Notas de Matem\'atica [Mathematical Notes], 90.

\bibitem{C}
\leavevmode\vrule height 2pt depth -1.6pt width 23pt, {\em Elementary
  introduction to new generalized functions}, vol.~113 of North-Holland
  Mathematics Studies, North-Holland Publishing Co., Amsterdam, 1985.
\newblock Notes on Pure Mathematics, 103.

\bibitem{GF}
{\sc G.~Fischer}, {\em Lineare {A}lgebra}, vol.~17 of Grundkurs Mathematik,
  Friedr. Vieweg \& Sohn, Braunschweig, fifth~ed., 1979.
\newblock In collaboration with Richard Schimpl.

\bibitem{FL1}
{\sc F.~G. Friedlander}, {\em The wave equation on a curved space-time},
  Cambridge University Press, Cambridge, 1975.
\newblock Cambridge Monographs on Mathematical Physics, No. 2.

\bibitem{Bible}
{\sc M.~Grosser, M.~Kunzinger, M.~Oberguggenberger, and R.~Steinbauer}, {\em
  Geometric theory of generalized functions with applications to general
  relativity}, vol.~537 of Mathematics and its Applications, Kluwer Academic
  Publishers, Dordrecht, 2001.

\bibitem{GlobTh}
{\sc M.~Grosser, M.~Kunzinger, R.~Steinbauer, and J.~A. Vickers}, {\em A global
  theory of algebras of generalized functions}, Adv. Math., 166 (2002),
  pp.~50--72.

\bibitem{HE}
{\sc S.~W. Hawking and G.~F.~R. Ellis}, {\em The large scale structure of
  space-time}, Cambridge University Press, London, 1973.
\newblock Cambridge Monographs on Mathematical Physics, No. 1.

\bibitem{HS}
{\sc J.~M. Heinzle and R.~Steinbauer}, {\em Remarks on the distributional
  {S}chwarzschild geometry}, J. Math. Phys., 43 (2002), pp.~1493--1508.

\bibitem{MOHor}
{\sc G.~H{\"o}rmann and M.~Oberguggenberger}, {\em Elliptic regularity and
  solvability for partial differential equations with {C}olombeau
  coefficients}, Electr. Jour. Diff. Equ.,,  (2004), pp.~1--30.

\bibitem{Hungerford}
{\sc T.~W. Hungerford}, {\em Algebra}, Hgolt, Rinehart and Winston, Inc., New
  York, 1974.

\bibitem{KKo1}
{\sc S.~Konjik and M.~Kunzinger}, {\em Generalized group actions in a global
  setting}, J. Math. Anal. Appl., 322 (2006), pp.~420--436.

\bibitem{KU2}
{\sc M.~Kunzinger}, {\em Generalized functions valued in a smooth manifold},
  Monatsh. Math., 137 (2002), pp.~31--49.

\bibitem{K}
\leavevmode\vrule height 2pt depth -1.6pt width 23pt, {\em Nonsmooth
  differential geometry and algebras of generalized functions}, J. Math. Anal.
  Appl., 297 (2004), pp.~456--471.
\newblock Special issue dedicated to John Horv\'ath.

\bibitem{3MikesV}
{\sc M.~Kunzinger, M.~Oberguggenberger, R.~Steinbauer, and J.~A. Vickers}, {\em
  Generalized flows and singular {ODE}s on differentiable manifolds}, Acta
  Appl. Math., 80 (2004), pp.~221--241.

\bibitem{PenroseKS}
{\sc M.~Kunzinger and R.~Steinbauer}, {\em A note on the {P}enrose junction
  conditions}, Classical Quantum Gravity, 16 (1999), pp.~1255--1264.

\bibitem{KS}
\leavevmode\vrule height 2pt depth -1.6pt width 23pt, {\em Foundations of a
  nonlinear distributional geometry}, Acta Appl. Math., 71 (2002),
  pp.~179--206.

\bibitem{KS1}
\leavevmode\vrule height 2pt depth -1.6pt width 23pt, {\em Generalized
  pseudo-{R}iemannian geometry}, Trans. Amer. Math. Soc., 354 (2002),
  pp.~4179--4199 (electronic).

\bibitem{KSV}
{\sc M.~Kunzinger, R.~Steinbauer, and J.~A. Vickers}, {\em Intrinsic
  characterization of manifold-valued generalized functions}, Proc. London
  Math. Soc. (3), 87 (2003), pp.~451--470.

\bibitem{GenConKSV}
\leavevmode\vrule height 2pt depth -1.6pt width 23pt, {\em Generalised
  connections and curvature}, Math. Proc. Cambridge Philos. Soc., 139 (2005),
  pp.~497--521.

\bibitem{Eyb4}
{\sc E.~Mayerhofer}, {\em On the characterization of p-adic
  {C}olombeau-{E}gorov generalized functions by their point values}, Math.
  Nachr., 280 (2007), pp.~1297-1301.

\bibitem{MObook}
{\sc M.~Oberguggenberger}, {\em Multiplication of distributions and
  applications to partial differential equations}, vol.~259 of Pitman Research
  Notes in Mathematics Series, Longman Scientific \& Technical, Harlow, 1992.

\bibitem{MO1}
{\sc M.~Oberguggenberger and M.~Kunzinger}, {\em Characterization of
  {C}olombeau generalized functions by their pointvalues}, Math. Nachr., 203
  (1999), pp.~147--157.

\bibitem{PSMO}
{\sc M.~Oberguggenberger, S.~Pilipovi{\'c}, and D.~Scarpal{\'e}zos}, {\em
  Positivity and positive definiteness in generalized function algebras}, J.
  Math. Anal. Appl., 328 (2007), pp.~1321--1335.

\bibitem{ON}
{\sc B.~O'Neill}, {\em Semi-{R}iemannian geometry}, vol.~103 of Pure and
  Applied Mathematics, Academic Press Inc., New York, 1983.
\newblock With applications to relativity.

\bibitem{Schw1}
{\sc L.~Schwartz}, {\em Sur l'impossibilit\'e de la multiplication des
  distributions}, C. R. Acad. Sci. Paris, 239 (1954), pp.~847--848.

\bibitem{SJ}
{\sc G.~W. Stewart and J.~G. Sun}, {\em Matrix perturbation theory}, Computer
  Science and Scientific Computing, Academic Press Inc., Boston, MA, 1990.

\bibitem{VW}
{\sc J.~A. Vickers and J.~P. Wilson}, {\em Generalized hyperbolicity in conical
  spacetimes}, Class. Quantum Grav., 17 (2000), pp.~1333--1260.

\end{thebibliography}

\end{document}